\documentclass{aa}
\usepackage{rotating}
\usepackage{natbib}
\usepackage{lscape}
\usepackage{color}
\usepackage{graphicx}

\newcommand{\mearth}{\mbox{$M_\oplus$}}
\newcommand{\mjup}{\mbox{$M_{Jup}$}}

\newcommand{\mic}{\mbox{$\mu$m}}
\newcommand{\app}{\mbox{$\sim$}}

\newcommand{\pp}{\mbox{$\pm$}}

\newcommand{\dra}{\mbox{$\Delta RA$}}
\newcommand{\ddec}{\mbox{$\Delta DEC$}}

\newcommand{\dg}{\mbox{$^\circ$}}
\newcommand{\prim}{\mbox{HR~4796~A}}

\newcommand{\eg}{e.g.}

\begin{document}

\title{The Gemini NICI Planet-Finding Campaign:\\ The Offset Ring of HR~4796~A
\thanks{Based on observations obtained at the Gemini
  Observatory, which is operated by the Association of Universities for
 Research in Astronomy, Inc., under a cooperative agreement with the
  NSF on behalf of the Gemini partnership: the National Science
  Foundation (United States), the Science and Technology Facilities
  Council (United Kingdom), the National Research Council (Canada),
  CONICYT (Chile), the Australian Research Council (Australia),
  Minist\'{e}rio da Ci\^{e}ncia e Tecnologia (Brazil) and Ministerio de
  Ciencia, Tecnolog\'{i}a e Innovaci\'{o}n Productiva (Argentina).}
}

\institute{European Southern Observatory, Alonso de Cordova 3107,  Vitacura, Casilla 19001, Santiago, Chile\label{inst1}\and
Institute for Astronomy, University of Hawaii, 2680 Woodlawn Drive, Honolulu, HI 96822, USA\label{inst2}\and
Institute for Astronomy, The University of Edinburgh, Royal Observatory, Blackford Hill, Edinburgh EH9 3HJ, United Kingdom\label{inst3}\and
Gemini Observatory, Southern Operations Center, c/o  AURA, Casilla 603, La Serena, Chile\label{inst4}\and
NASA Goddard Space Flight Center, Exoplanets and Stellar Astrophysics Laboratory, Greenbelt, MD 20771\label{inst5}\and
Steward Observatory, University of Arizona, 933 North Cherry Avenue, Tucson, AZ 85721\label{inst6}\and 
Mauna Kea Infrared, LLC, 21 Pookela St., Hilo, HI 96720\label{inst7}}

\author{Zahed Wahhaj\inst{\ref{inst1}}\and 
Michael C. Liu\inst{\ref{inst2}}\and  
Beth A. Biller\inst{\ref{inst3}}\and
Eric L. Nielsen\inst{\ref{inst2}}\and
Thomas L. Hayward\inst{\ref{inst4}}\and
Marc Kuchner\inst{\ref{inst5}}\and
Laird M. Close\inst{\ref{inst6}}\and 
Mark Chun\inst{\ref{inst2}}\and
Christ Ftaclas\inst{\ref{inst2}}\and
Douglas W. Toomey\inst{\ref{inst7}}.
}

\date{25 December 2013}

\abstract{We present $J, H, CH_{4}$ short (1.578~\mic), $CH_{4}$ long
  (1.652~\mic) and $K_s$-band images of the dust ring around the
  10~Myr old star \prim\  obtained using the Near Infrared Coronagraphic Imager (NICI) on the 
  Gemini-South  8.1~meter Telescope.
  Our  images clearly show for the first time the
  position of the star relative to its circumstellar ring thanks to
  NICI's translucent focal plane occulting mask.
  We employ a Bayesian Markov Chain Monte Carlo method to constrain the offset vector between
  the two. The resulting probability distribution shows that the ring center is offset 
  from the star by 16.7\pp1.3~milliarcseconds along a position angle
  of \textcolor{black}{26\pp3\dg , along the PA of the ring, 26.47\pp0.04\dg }. We find that the
  size of this offset is not large enough to explain the brightness
  asymmetry of the ring. The ring is measured to have mostly red reflectivity
  across the $JHK_s$ filters, which seems to indicate \textcolor{black}{micron-sized} grains.
  Just like Neptune's 3:2 and 2:1 mean-motion resonances delineate the inner and outer edges 
  of the classical Kuiper Belt, we find that the radial extent of the \prim\ and the
  Fomalhaut rings could correspond to the 3:2 and 2:1 mean-motion
  resonances of hypothetical planets at 54.7~AU and 97.7~AU in the two systems, 
  respectively.  \textcolor{black}{A planet orbiting \prim\ at 54.7~AU would have to be
  less massive than 1.6~\mjup\ so as not to widen the ring too much by
  stirring.} \\
\\
 {\bf Accepted by A\&A on 23 April 2014.}
}


\maketitle

\section {INTRODUCTION}

Debris disks are composed of dust produced by collisions between
planetesimals orbiting stars ($\gtrsim$~10~Myr)
\citep[\eg][]{1993prpl.conf.1253B,2008ARA&A..46..339W}. 
Since the first image of a debris disk around $\beta$ 
Pictoris \citep{1984Sci...226.1421S}, more than three dozen debris 
disks have been resolved in the optical, infrared and submillimeter\footnotemark. Most of these disks
exhibit asymmetries which are difficult to explain without invoking a
dynamical perturber, for example a planet \citep{2008ARA&A..46..339W}.
In the case of $\beta$ Pictoris, there is now evidence that the 
recently discovered planet \citep{2009A&A...493L..21L,
  2010Sci...329...57L}, may be directly responsible for one of these 
asymmetries, a warp in the disk
\citep{2012A&A...542A..40L,2000ApJ...539..435H,1997MNRAS.292..896M}.
Moreover, many of the directly imaged planets 
have been found around A stars with debris disks
\citep{2010Natur.468.1080M, 2010Sci...329...57L, 2013ApJ...775...56K,
  2013ApJ...779L..26R}, which suggests that these systems at one time 
possessed massive primordial disks. Thus debris disks may indicate the 
presence of planets in two ways: (1) by their morphology and (2) by their
presence alone.

\footnotetext[1]{The Catalog of Circumstellar Disks: http://www.circumstellardisks.org/}    

The debris ring around the young (8-10~Myr;
\citealt{1995ApJ...454..910S}) A0~star HR~4796A has two 
asymmetries: (1) a brightening of its North-East ansa with respect 
to its South-West ansa
\citep{1998ApJ...503L..83K,1999ApJ...527..918W,1999ApJ...513L.127S},
and (2) a possible offset of the ring center from the star \citep{2009AJ....137...53S,2011ApJ...743L...6T}.
\citet{1999ApJ...527..918W} have argued that these two phenomenon may 
be related to the greater exposure of the ring to stellar radiation at the pericenter.  
However, the offset may also be caused by the 
stellar companion \textcolor{black}{\citep[M2.5V;][]{2006A&A...459..511B}} at 7.86$''$ (Epoch UT April 6, 2012)
which has a position angle (PA) similar (45\dg\pp 1$+$180\dg) to that
of the ring \citep{1993ApJ...418L..37J}. The ring has a PA of 26.01\pp0.16\dg\ and an 
inclination to the line of sight of 75.88\pp0.16\dg\ \textcolor{black}{ \citep[$\approx$ 14\dg\ from
edge-on;][]{2009AJ....137...53S}}. It has been suggested that the narrowness of the ring could
be due to inner and outer \textcolor{black}{planetary} companions \citep{1999ApJ...527..918W}. Any 
sharp outer cutoff cannot be explained by the stellar companion
alone \citep{2010A&A...524A..13T}. On the other hand, it appears that
the outer edge of the ring may have a long tenuous tail 
\citep{2005ApJ...618..385W,2011ApJ...743L...6T}. \textcolor{black}{However, the narrow
``streamers'' emanating from the outer edges of the ring as reported
in \citet{2011ApJ...743L...6T}  appear to be an artifact from image
processing that is only significant in angular differential observations
with insufficient total sky rotation \citep{2012A&A...542A..40L,2012A&A...545A.111M}.} 
\citet{2005ApJ...618..385W}
found that combined modeling of the Keck mid-infrared (MIR) images and HST near-infrared (NIR)
images required the existence of a wide ring (extending from 0.7$''$
to 1.9$''$) which was 10 times more tenuous than the narrow ring with
radius 1.05$''$ \textcolor{black}{\citep[76.4~AU for a distance of 72.78~pc;][]{2007A&A...474..653V}}.   
The wide ring, with smaller and lower albedo grains than the
narrow ring, could be interpreted as dust being blown away by
radiation pressure.  \textcolor{black}{The source of the blow-out dust 
would mainly be planetesimals in the narrow ring, 
but also could include a lower density population interior to it.}

In this paper, we present images of the HR~4796A ring taken with the Near Infrared
Coronagraphic Imager (NICI) at the 8-m Gemini South Telescope in the $JHK_s$ and two methane bands.
These images clearly show, for the first time, the positions of the
star and the ring so that any relative shift can be measured very precisely. The ring ansae asymmetry is also detected in 
all five bands, and thus confirmed unambiguously here. Lastly, we
review the possible causes of these two asymmetries.

\section{OBSERVATION}
We observed HR~4796A on UT 2009 January 14 and UT 2012 April 6 \& 7
with NICI \citep{2008SPIE.7015E..49C} as part of the Gemini NICI
Planet-Finding Campaign on the Gemini-South 8.1~m Telescope \citep{2010SPIE.7736E..53L,2013ApJ...773..179W,2013ApJ...776....4N,2013ApJ...777..160B}. The star was
observed in Angular Difference Imaging
(ADI; \citealt{2004Sci...305.1442L,2006ApJ...641..556M}) mode, with two different
filters in the NICI's two cameras. On 
2009 January 14, we used the $CH_4S$ ($\lambda$=1.578~\mic) and $CH_4L$
($\lambda$=1.652~\mic) moderate-bandwidth ($\Delta\lambda$/$\lambda$=4\%)
filters with a 50/50 beamsplitter sending light to the two
cameras \textcolor{black}{(plate scales 17.96\pp 0.01~mas/pixel for $CH_4S$  and
17.94\pp 0.01~mas/pixel for $CH_4L$)}, which were read out simultaneously. The integration time per image was 1 minute. 
The star was placed behind the semi-transparent (0.28\% central transmission) focal plane
mask with a half-transmission radius of 0.32$'"$. The central
opacities for the various filters are provided in
Table~\ref{tab:maf}. These were measured by observing a pair of stars
4.2$''$ apart of known flux ratio from the 2MASS catalog, as described
in \citet{2011ApJ...729..139W}.

\textcolor{black}{Thanks to the focal plane mask}, \prim\ was imaged without saturation so
that its position with respect to the debris ring can be measured very
precisely. On 2012 April 6, we imaged simultaneously in the $H$ and
$K_s$ bands and on April 7, we imaged only in $J$ band, sending all the light to
one camera by replacing the dichroic with a mirror. \textcolor{black}{In the April 6 observations, the
star was only lightly saturated, exhibiting no
obvious change in the point-spread-function (PSF) shape.} The total
number of images taken and the total sky rotation obtained in ADI mode
\textcolor{black}{at each epoch} are presented in Table~\ref{tab:obs}. \textcolor{black}{The
smearing of the sky due to ADI mode during an individual image} was
$<$1.5\dg\ for all observations, or $<$0.5 FWHM of the NICI PSF at 1$''$.

\section{DATA REDUCTION}

We aim to estimate the center of the \prim\ ring and compare it to the location of the star. 
For this purpose we focus only on the brightest parts of the ring. The brightness decreases
quickly on either side of the bright rim.
Thus, we use the point-source recovery ADI pipeline as described in
\citet{2013ApJ...779...80W}, since it is most successful at removing starlight and isolating the brightest parts of the ring.
The steps of the pipeline are:

\begin {description}
\item{1.} Do basic reduction: apply flatfield, distortion and position angle corrections.
\item{2.} Find centroids and apply image filters to the images.
\item{3.} Subtract the median of the stack from the individual
  difference images (ADI subtraction).
\item{4.} De-rotate and stack the images. 
\end {description}

In this pipeline, we filter the images to remove emission at spatial scales that are less than half
or more than twice the FWHM of the \textcolor{black}{NICI $H$-band PSF ($\approx$ 3 pixels or 54~mas)}. Thus ring features which are much 
larger or smaller than the NICI PSF are not present in the reduced image. 

In step 3 of the pipeline, the median image is fit to each science image
to minimize the RMS in an annulus in the difference image. 
The fit is performed using a simplex-downhill method which searches
for optimum intensity scalings and horizontal and vertical shifts. 
The annulus used for the fit has inner and outer radii of 0.6$''$ and
0.8$''$, respectively.
 
\begin{figure*}[ht]
    \hbox{
      \includegraphics[width=20cm]{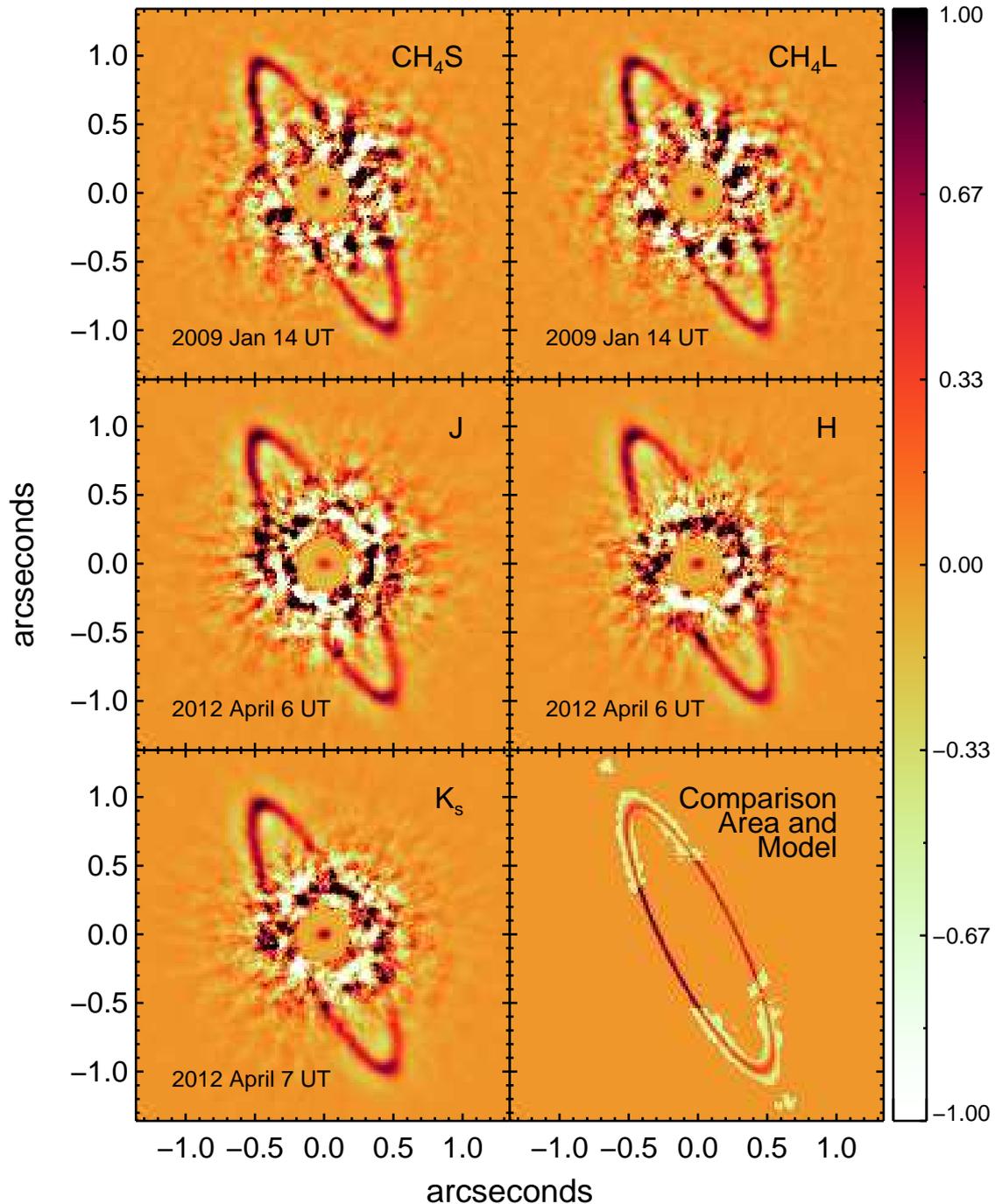}
    }
\caption{Unsaturated images of HR 4796A, with both the star and dust ring detected with high astrometric precision. 
  Reduced images from five filters taken in two epochs are shown. The black spot in the centers of the panels 
  show the star (unsaturated in $J$, $CH_4$ short and $CH_4$ long; see
  Table~2). \textcolor{black}{The intensities within 180 mas of the star have been
  divided by 100 to clearly show the unsaturated stellar peak}. The reductions were done using a point-source recovery pipeline, which 
  yields a higher fidelity image of the bright rim of the \prim\ disk. The last panel to the bottom right is the best-fit model 
  to the $K_s$-band image \textcolor{black}{with the model and data comparison region
  highlighted (by subtracting 0.5 from the region's pixel values). The
  color bar to the right gives the normalized pixel intensities in
  linear scale.}}
\label{fig:all_filts}
\end{figure*}

The subtraction of the reference PSF in step 3 causes different amounts of flux to
be lost from different parts of the ring, potentially changing its morphology. However, since we 
have 24--81\dg\ of sky rotation (see Table~\ref{tab:obs}), the image of the ring ansae (radius 1$''$) moves 
over more than 20 pixels on the detector as the science images are obtained.  As a result, the contribution from the 
ring to the median combination of the stack and thus to the PSF subtraction is negligible.
The reduced images are shown in Figure~\ref{fig:all_filts}.

\section{ANALYSIS}
\label{section:results}
We have reduced images in five filters of the bright rim of the \prim\ ring, two
obtained in January 2009 ($CH_4$ short and $CH_4$  long) 
and three in April 2012 ($JHK_s$). In these images, both the location of the star and 
the ring are captured at a resolution of 54~mas. In this section, we model the rim as a ring with a gaussian profile, viewed at 
some PA and inclination, with an offset of the ring center from the star.

We use three different methods to determine the offset of the ring
center \textcolor{black}{relative to the star}:
\begin {description}
\item{1.} Radial profiles along different PAs toward the NE and SW
  ansae to determine the maximum separations of the ring in the two
  directions (without a ring model).
\item{2.} Bayesian MCMC analysis to produce probability distributions
  for the ring offset in RA and DEC.
\item{3.} Bayesian MCMC analysis like method 2, but also modeling the self-subtraction of the ring.
\end {description}

\subsection {Method 1: Radial Profiles}

\begin{figure*}[ht]
  \includegraphics[width=16cm]{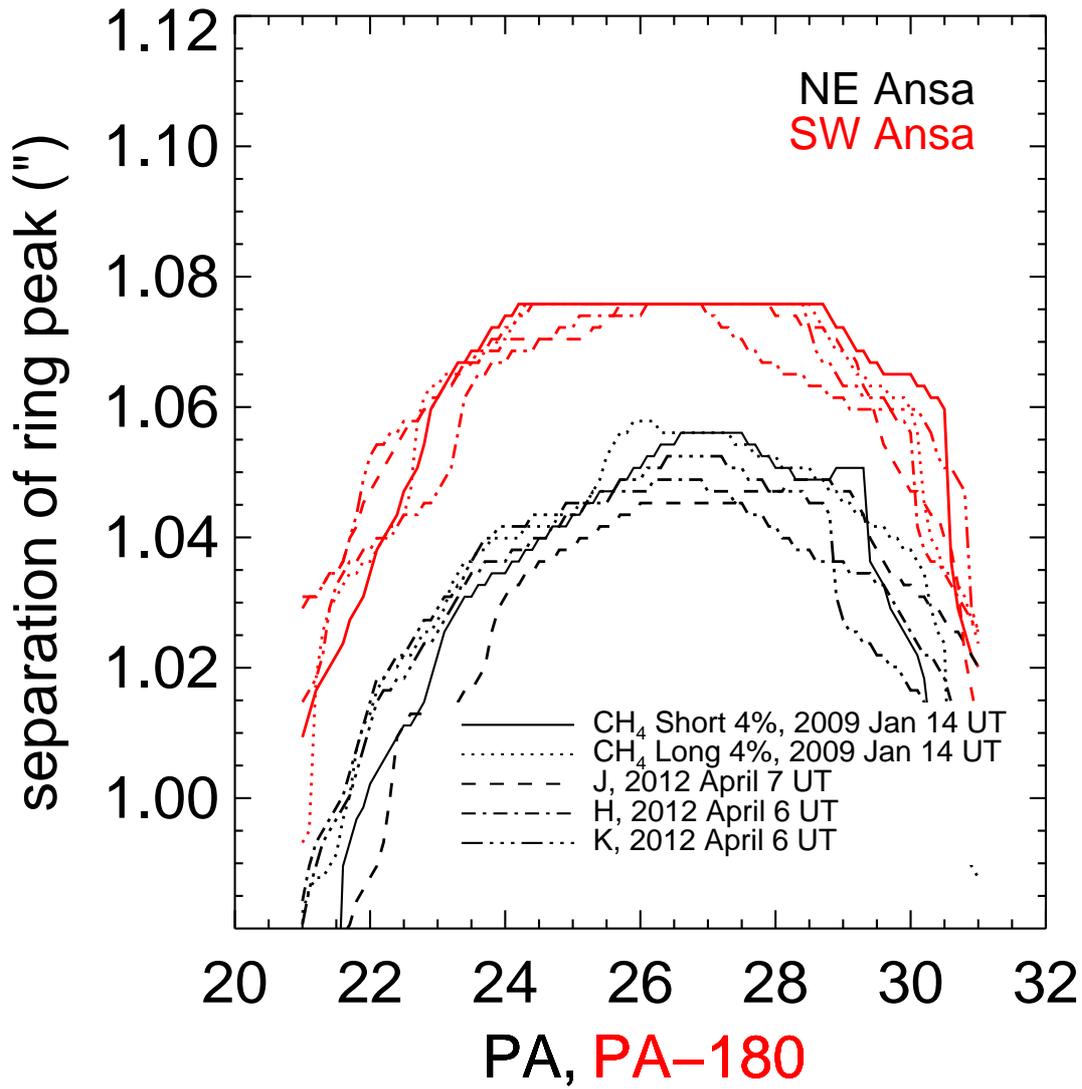}

  \caption{Line cuts taken from the stellar position along different PAs towards the ring ansae. The PA towards the SW ansa,
shown in red, has had 180\dg\ subtracted from it for easy comparison with the NE ansa (shown in black). This illustrates that the NE and SW peaks are not equally separated from the star. In other words, the ring is offset from the star.}
  \label{fig:linecuts}
\end{figure*}

We first study the radial profiles starting at the stellar position, pointed along several PAs towards the ring ansa,
at increments of 0.1\dg . \textcolor{black}{ Since the star is detected
  at very high signal to noise and is unsaturated, the uncertainty in
  its location is only 0.2~mas and is limited by the centroiding
  algorithm and the pixel size \citep{2013ApJ...779...80W}.} For each of these profiles, we estimate by cubic interpolation the projected separation of the brightest 
peak. Figure~\ref{fig:linecuts} shows these \textcolor{black}{projected separations} for the five filters. It is clear the NE ansae is consistently 
found to be closer to the star. Moreover, the peaks in the NE and SW ansae are found to lie along a straight line, giving credence 
to the method. 

\subsection {Method 2: Bayesian MCMC Fitting}

To quantify the ring offset, we compare a ring model to our NICI images using the $\chi^2$ statistic. 
\textcolor{black}{To avoid the effect of regions where the ring is faint compared to
residuals from the PSF subtraction,} we mask out regions not close to the ring ansae and exclude them from
our fitting. Regions within 0.55$''$ of the star 
are also not included in the comparisons. The ring is known to be inclined to the line of sight at an angle of 75.88$\pp$0.16\dg \citep{2009AJ....137...53S}.
We use this inclination to also exclude all regions with projected separations outside the 0.8$''$ to 1.5$''$ range. Furthermore, regions with negative 
emission which result from filtering the images are also excluded from
the comparison, since they are not physical. \textcolor{black}{The comparison region is
shown together with the model for the $K_s$-band in
Figure~\ref{fig:all_filts}. For the $K_s$-band, a total of 1041 pixels
or approximately 147 resolutions elements make up the comparison
region. The region has a similar size in the other bands.}  

For reasonable $\chi^2$ statistics, we need a good estimate of the noise in the images. In order to remove all 
spatial features much larger than two pixels, we convolve the data with a
Gaussian kernel of 2 pixel FWHM and subtract the result from the
data. We then take the standard deviation of the pixel values in the
fitting region \textcolor{black}{(defined above)} as the noise when 
calculating  $\chi^2$. This standard deviation is 8 times larger than
that of a Gaussian distribution with the same FWHM, indicating a
distribution with long tails. Indeed the ring emission is still contributing 
to the standard deviation as seen in the residual image. 
However, it is preferable to over estimate the noise since then the final
constraints we obtain will be  conservative. 
\\
\\
\\
The $\chi^2$ statistic is given by

\begin{equation}
  \chi^2 = \sum_{pixels} {{(Data-Model)^2} \over {noise^2}}
\end{equation}
 
The brightness of the ring $B(r)$ is modeled as a circular annulus of mean radius, $r_0$, with a radial brightness profile described by a gaussian of width $\sigma$
\begin{equation}
  B(r) = B_0 e^{ -(r_0-r)^2/{2 \sigma^2} }
\end{equation}

The additional parameters of the model are the inclination, PA, and
ring center offsets, \dra\ and \ddec , totalling seven parameters which completely describe the model.
As a starting point for the Bayesian MCMC computations, we find the best-fit parameters using the simplex-downhill IDL routine, {\it AMOEBA} \citep[\eg][]{1992nrca.book.....P}.

We use a Metropolis-Hastings Markov Chain Monte Carlo method to
calculate the posterior probabilities for the seven ring parameters. 
According to Bayes's theorem, the probability of a model given the data is
$$P(Model | Data) = P(Data | Model) \frac{P(Model)}{P(Data)}. $$  Since we do not
intend to include prior information about the data or the model, we set $P(Model) / P(Data) = 1$. 
The probability of the data given the model, $P(Data | Model)$, is given by $e^{-\chi^2/2}$.

We start at a given point in the seven-dimensional parameter space of models. Next, a trial jump to a new 
point in the parameter space is evaluated on the basis of its probability relative to the current point. Whether the jump to the 
new location is accepted is decided by the Metropolis-Hastings measure
\citep[\eg\ for a full description see][]{Gregory05}. If the trial jump is accepted, the new location becomes the current location.
Otherwise, the current location stays the same. In either case, a new trial jump is considered. This process is repeated until we converge 
to an equilibrium. The entire set of locations at each step gives the
posterior probability distribution for the seven model parameters\citep{1953JChPh..21.1087M}. 

The  trial jump is given by a seven-dimensional vector chosen randomly
from a seven-dimensional Gaussian distribution with 
appropriate standard deviations along each dimension. These standard
deviations are chosen so that the trial jump is accepted at a
rate between 25\% and 75\%. If the accepted rate is not within this range, then the solution will take too long to converge.

\begin{figure*}[ht]
    \vbox{
     \hbox {
        \includegraphics[width=8cm]{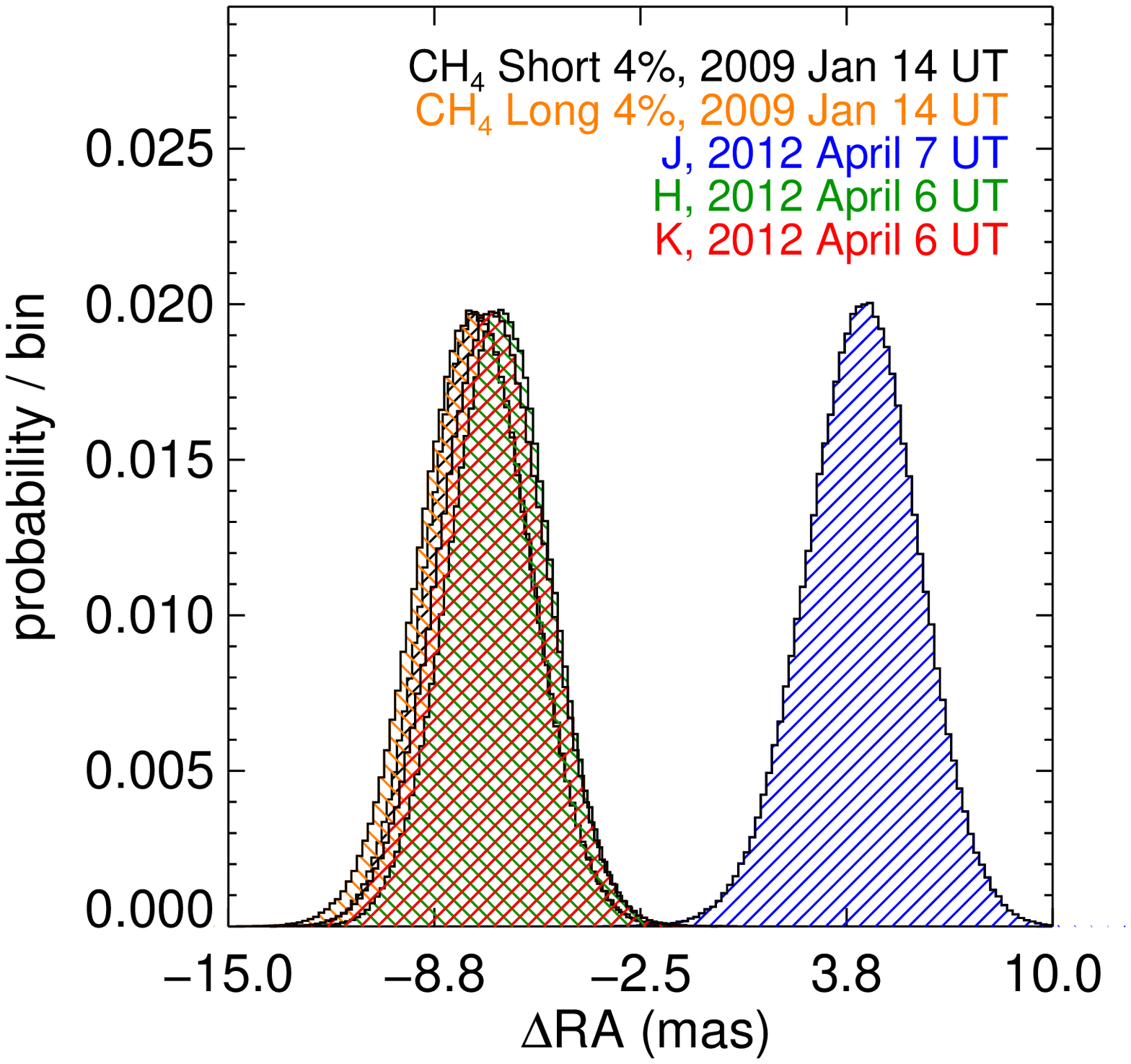}
        \includegraphics[width=8cm]{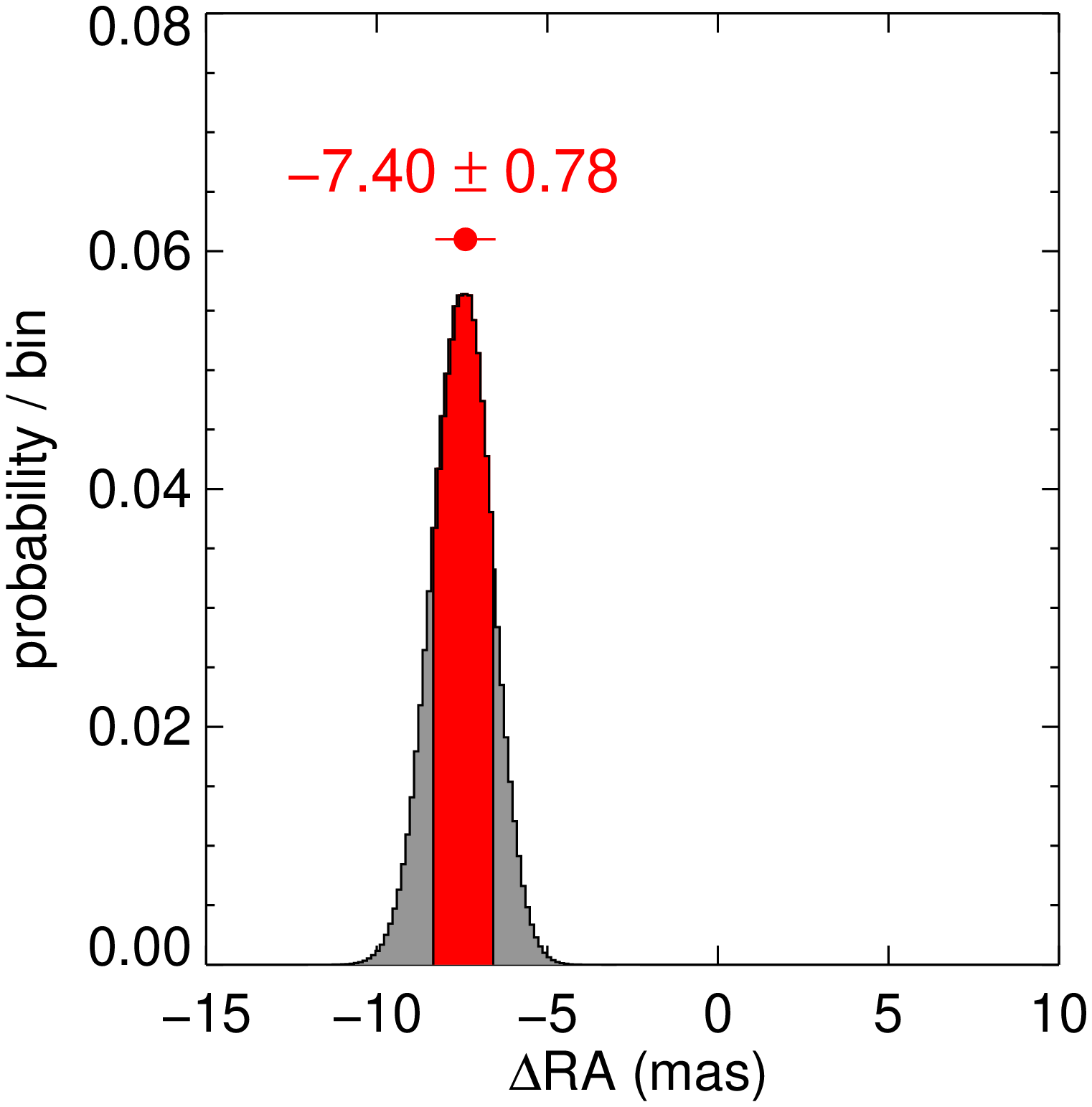}
      }  
      \hbox {
        \includegraphics[width=8cm]{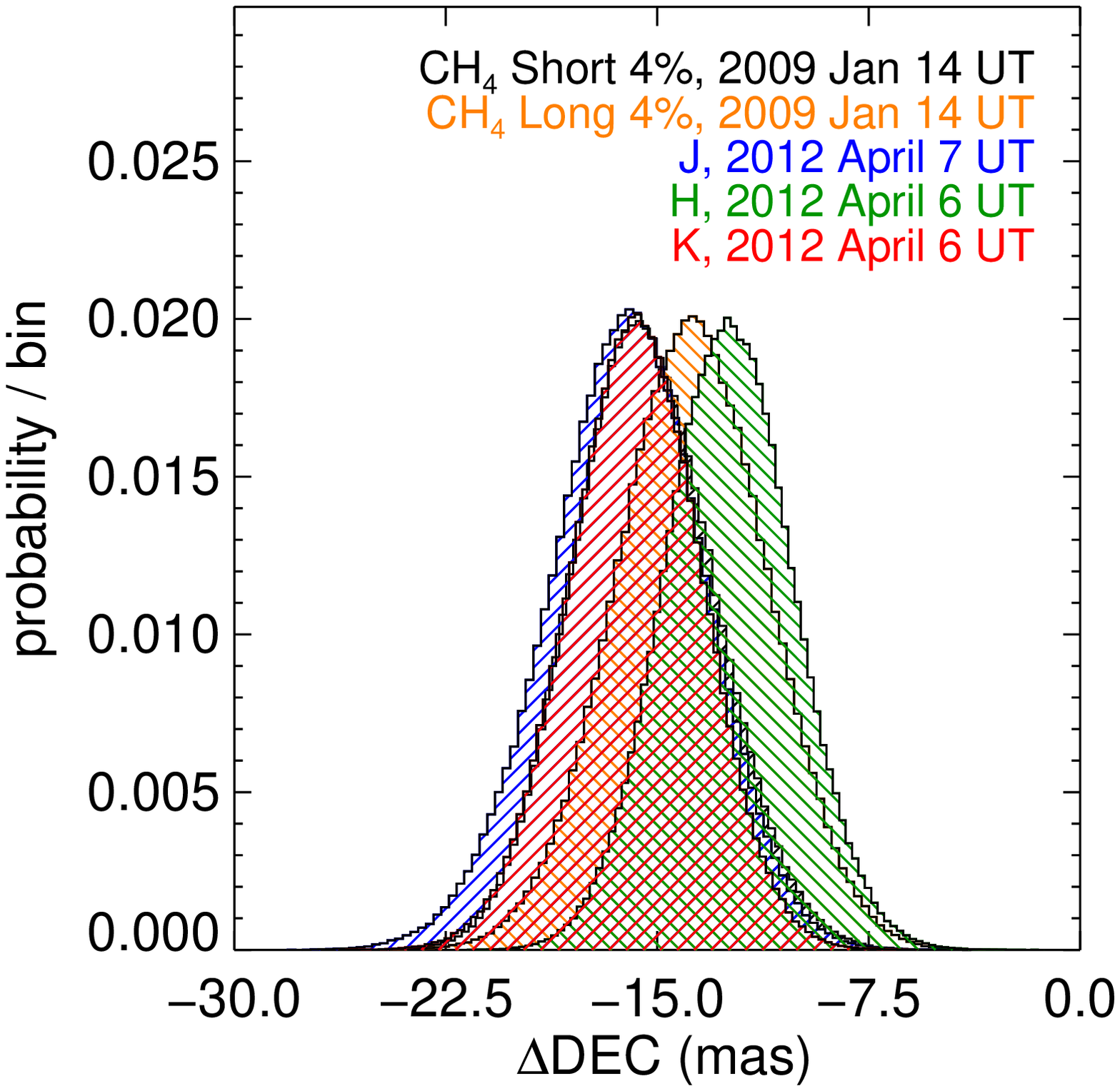}
        \includegraphics[width=8cm]{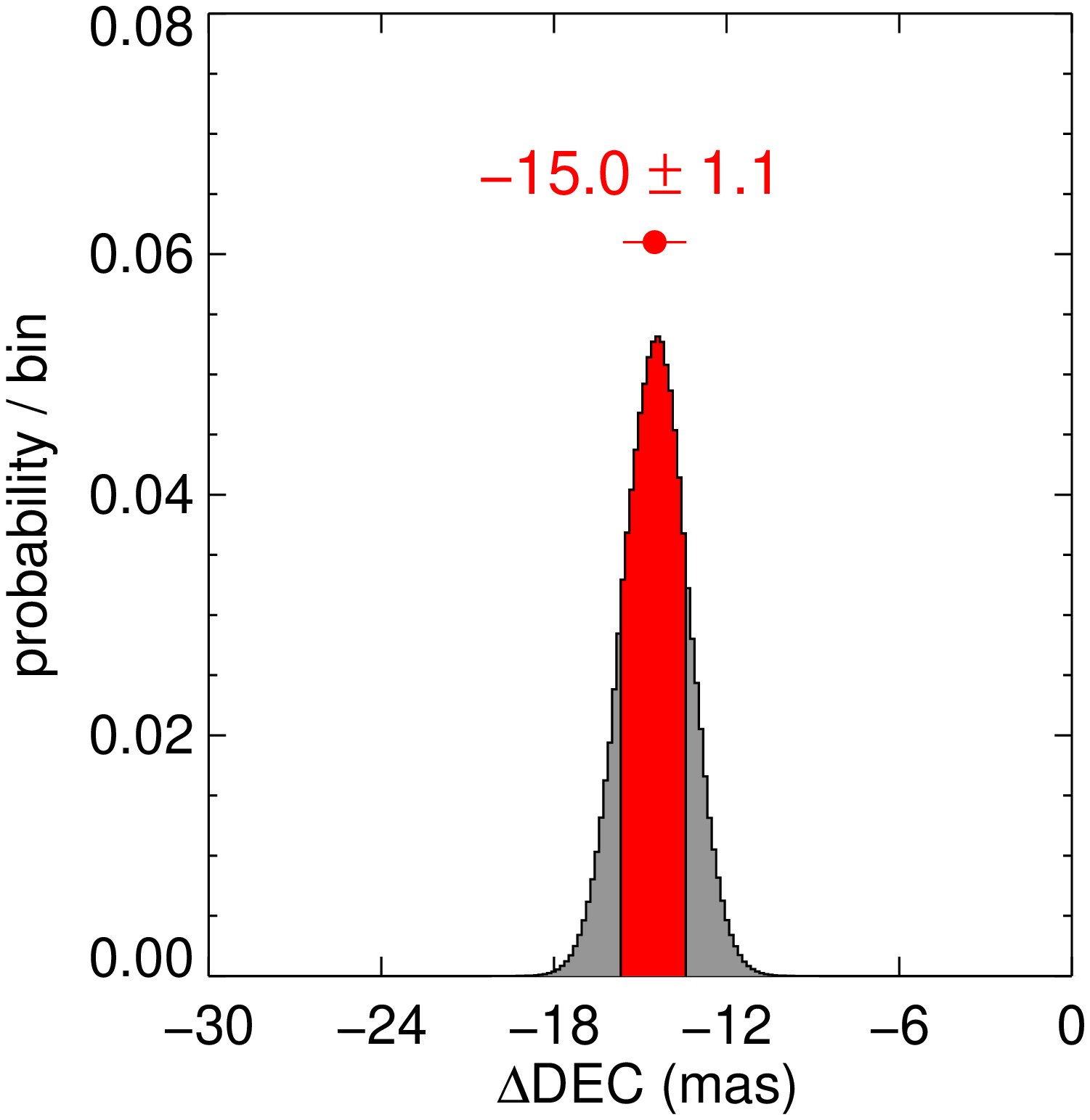}
      }}  
  \caption{Top left: Probability distribution for the RA offset of the
    \prim\ ring for each of the 5 filters. Right: The cumulative
    probability distribution \textcolor{black}{with the 1-$\sigma$
      dispersion in the distributions shown in red.} Bottom: The same for the DEC offset of
    the \prim\ ring. The offset of the ring from the stellar
    position in RA is in good agreement for all filters, except $J$-band,
    where it differs by half a NICI pixel (\app10 mas). 
    This is because the star was 3--5 pixels offset from the
    center of the mask during the $J$-band observations, which likely led to
    a systematic error in our centroid estimate. 
}
  \label{fig:radec_dists}
\end{figure*}

\begin{figure*}[ht]

  \centerline{
    \vbox{
     \hbox {
        \includegraphics[width=6cm]{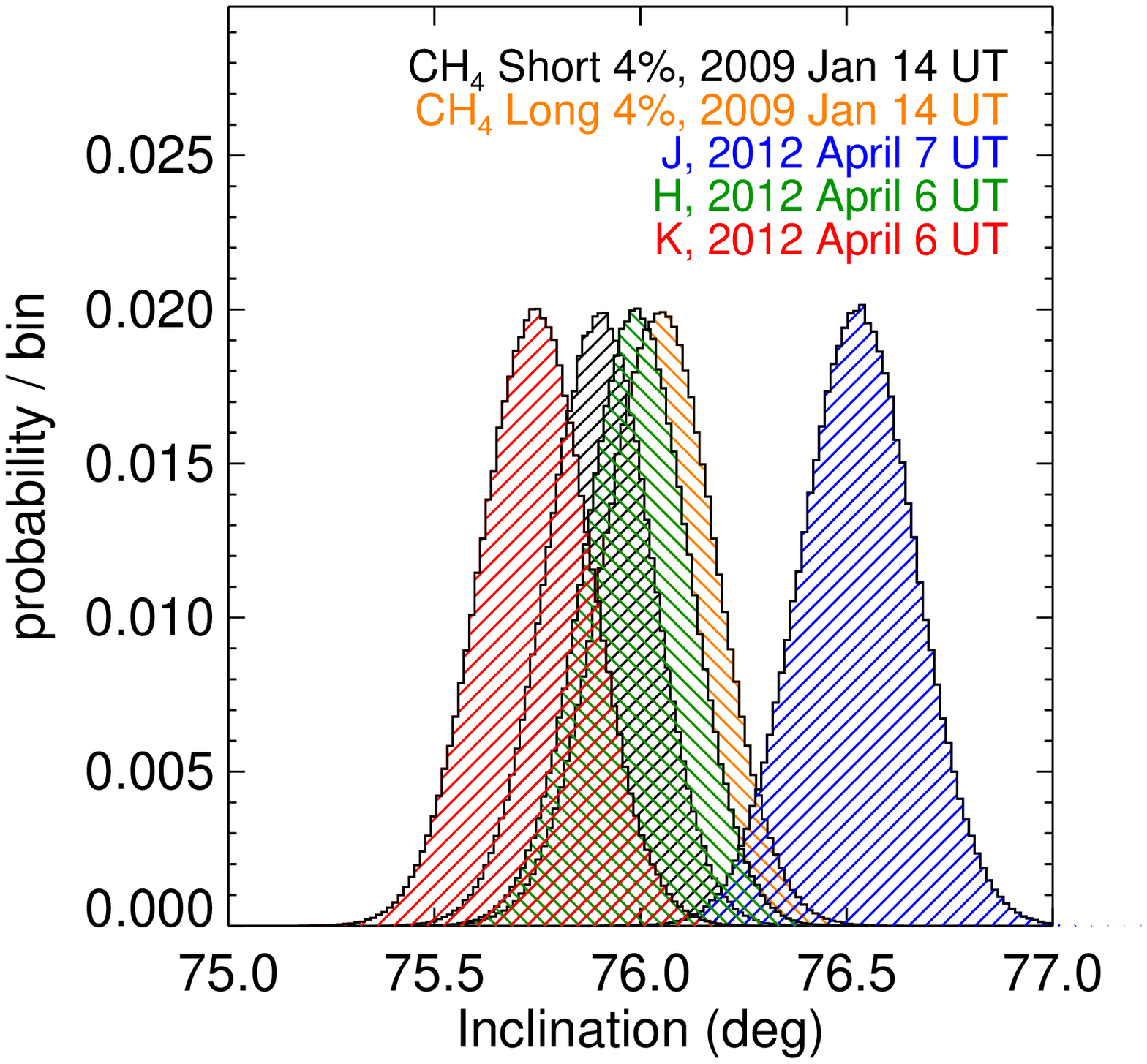}
        \includegraphics[width=6cm]{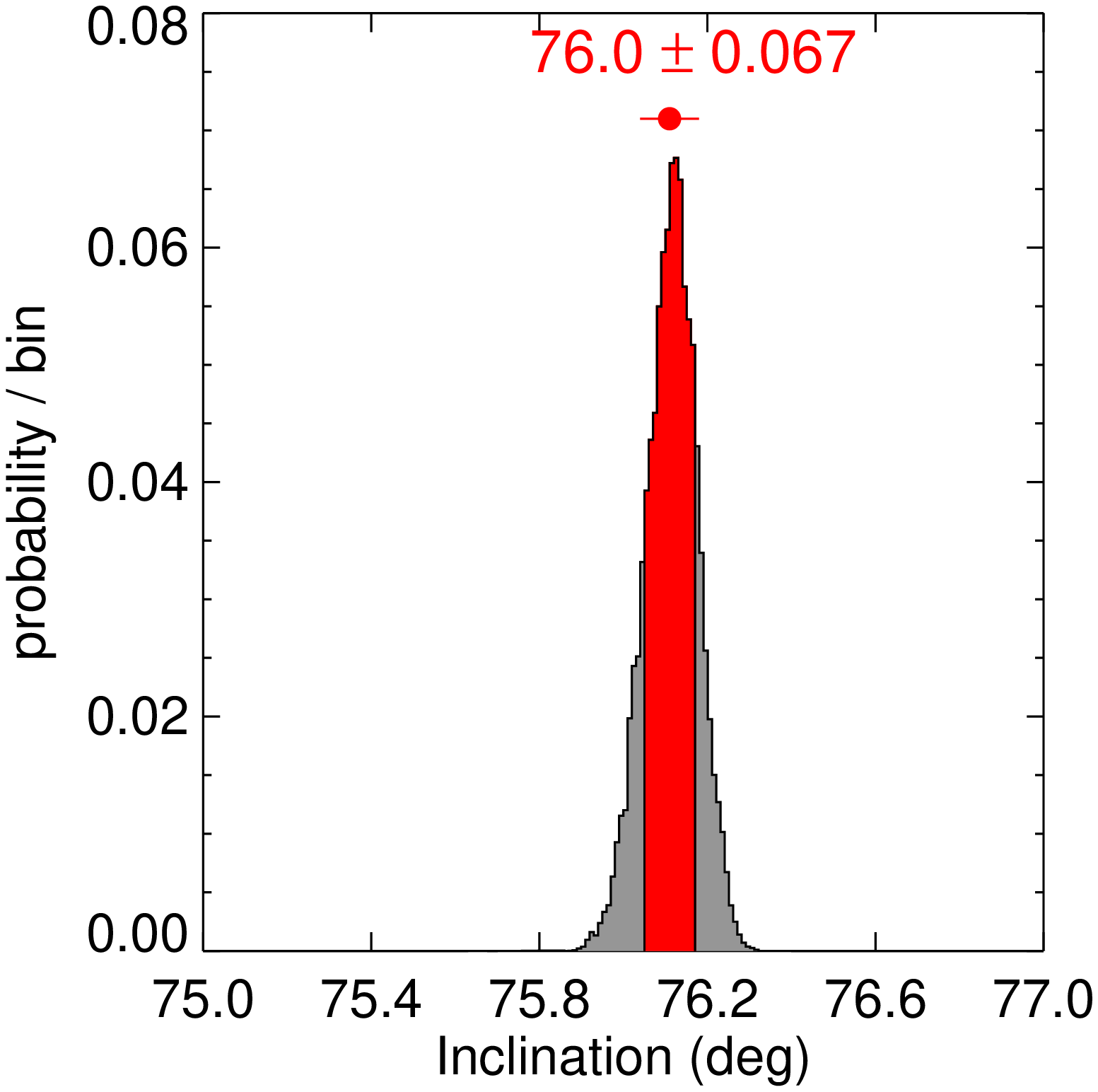}
      }  
      \hbox {
        \includegraphics[width=6cm]{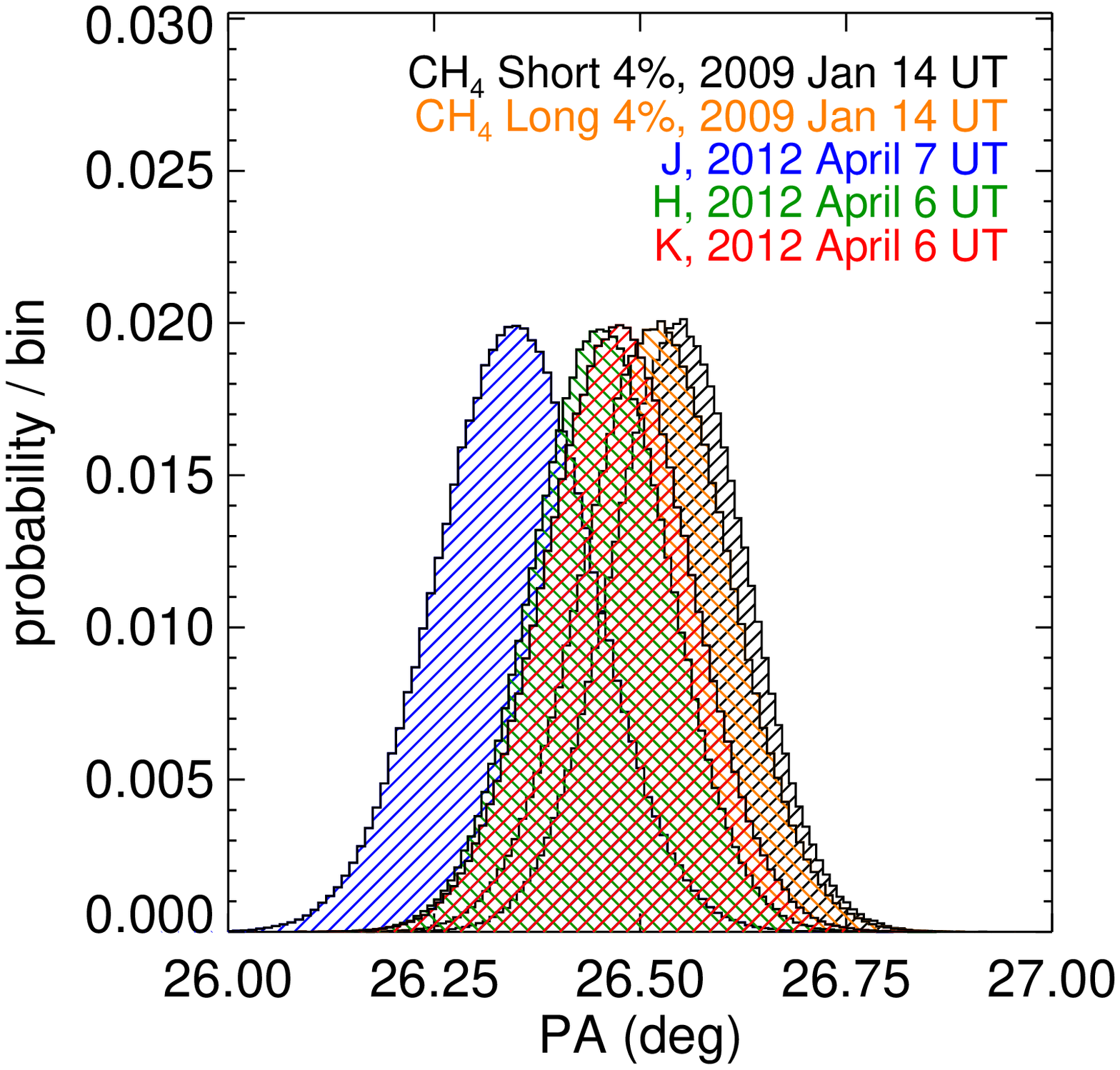}
        \includegraphics[width=6cm]{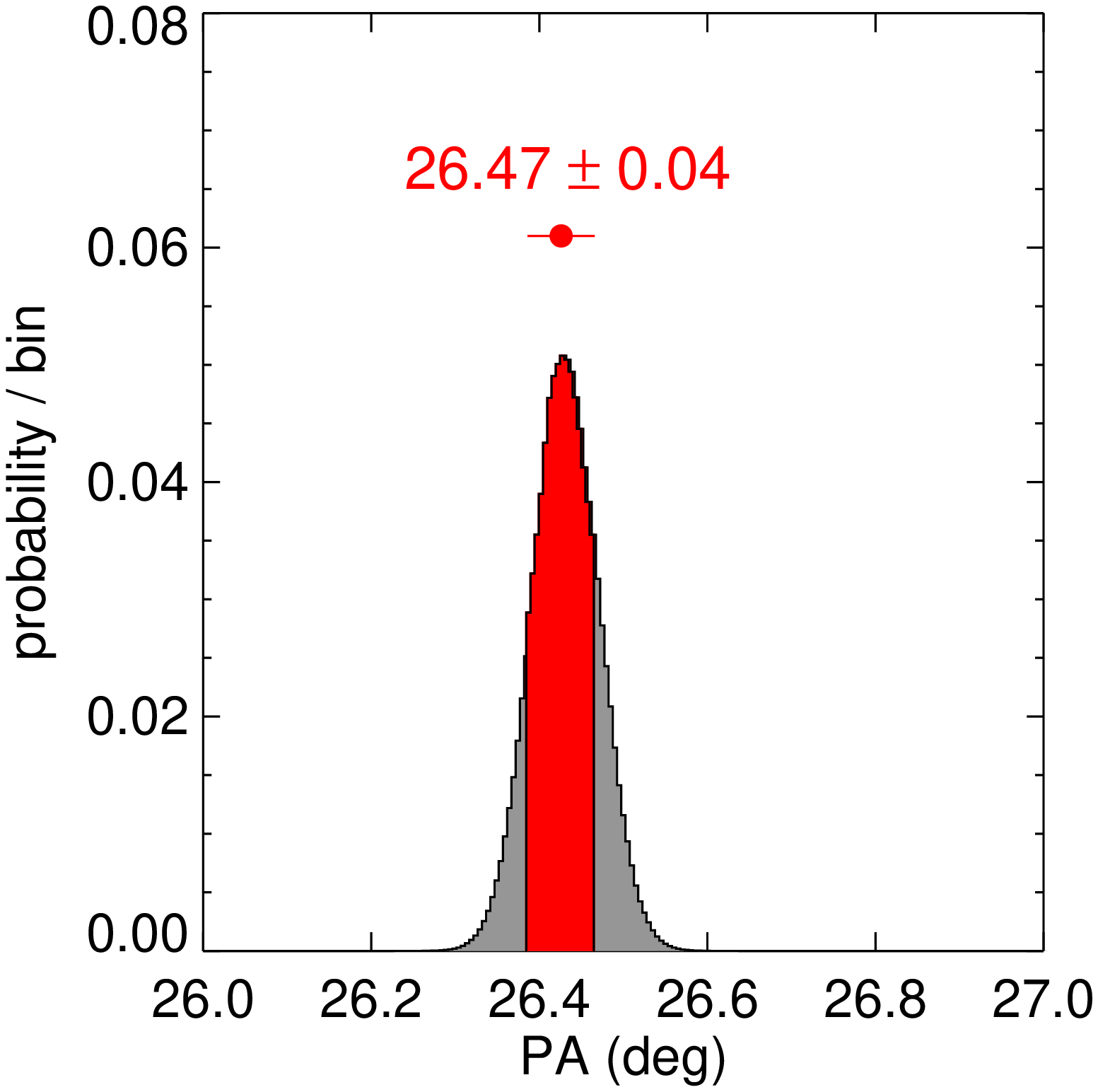}
      }
      \hbox {
        \includegraphics[width=6cm]{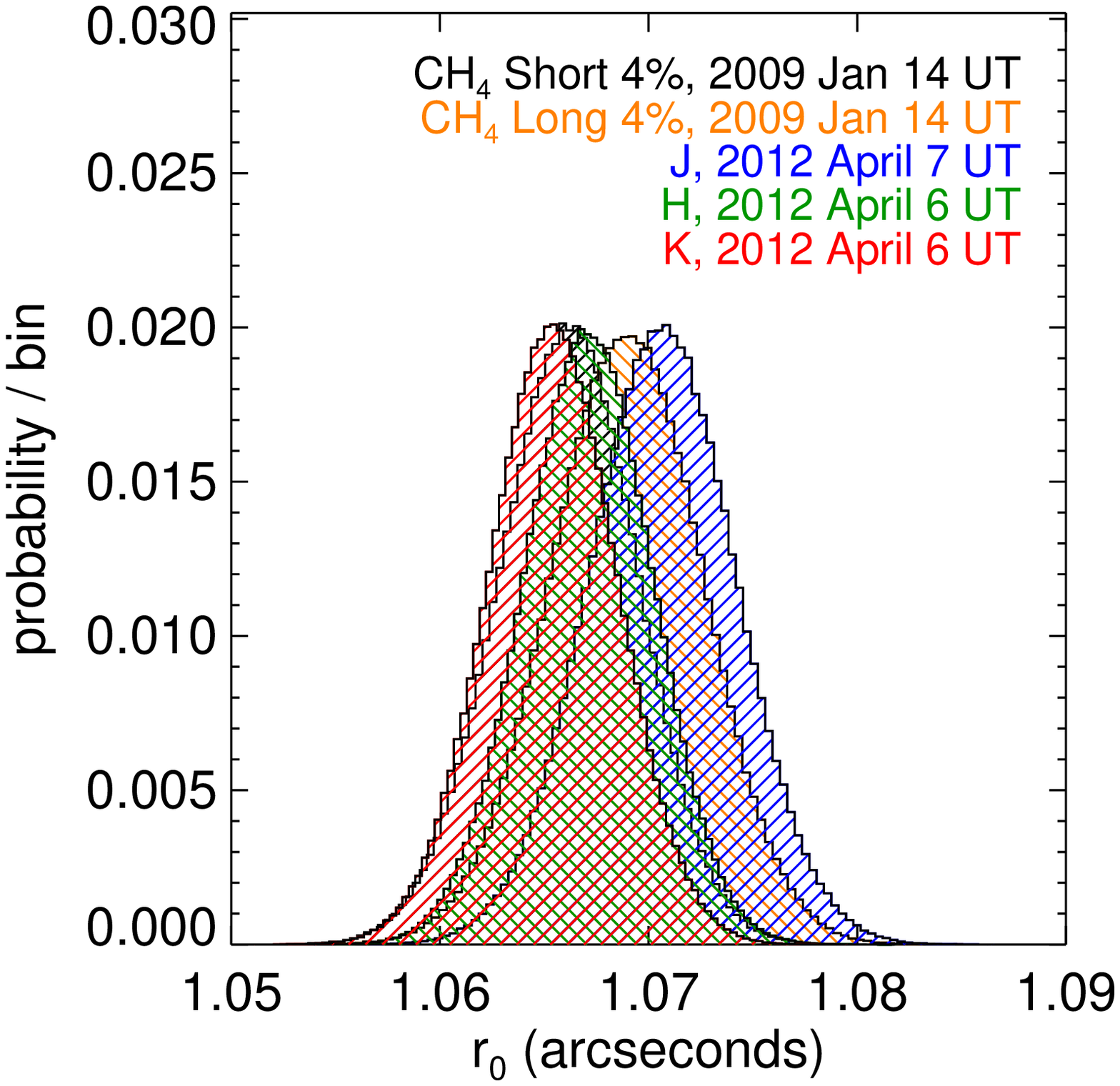}
        \includegraphics[width=6cm]{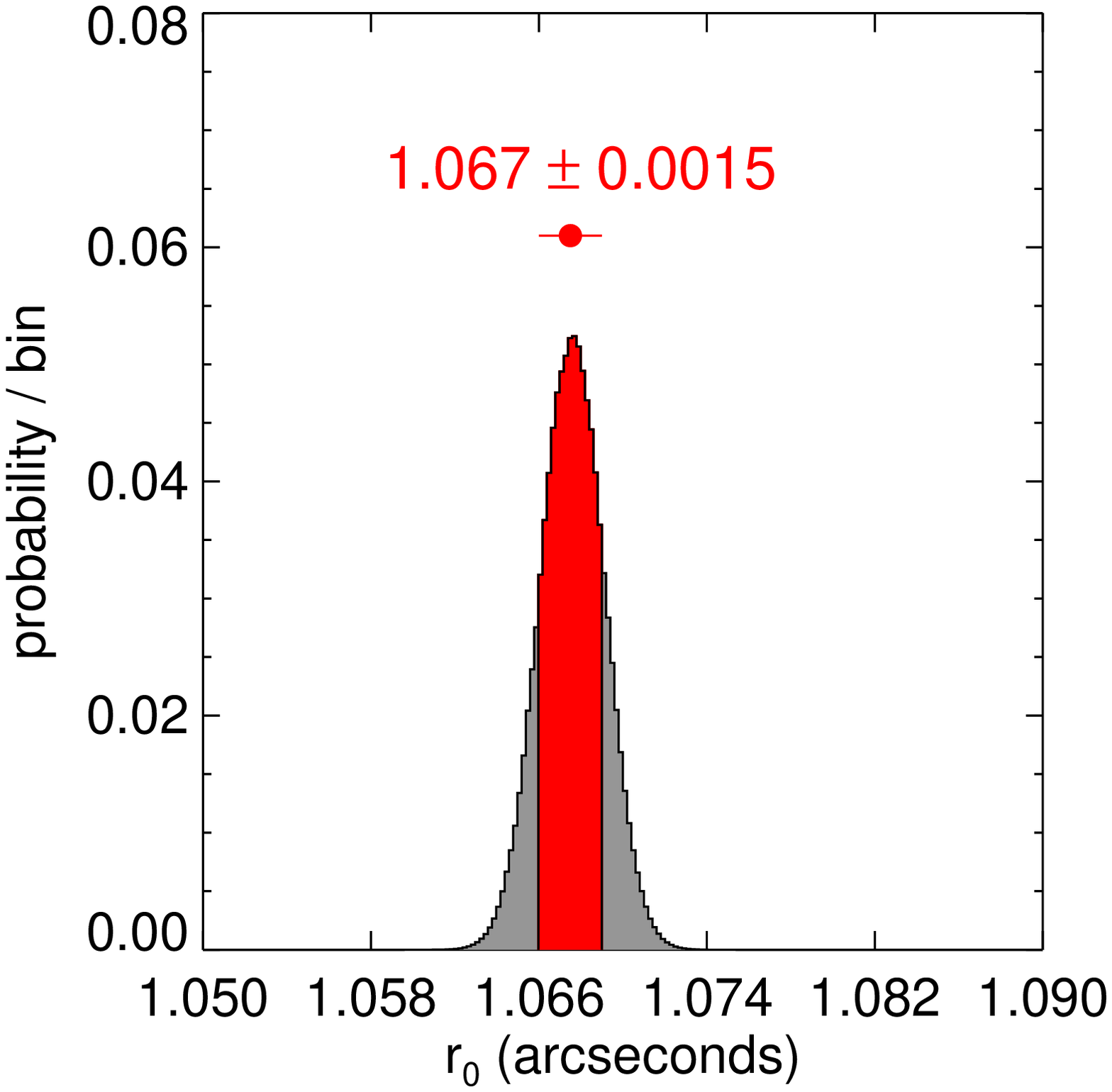}
      }}  
  }
  \caption{Probability distributions for the other ring parameters.}
  \label{fig:other_dists}
\end{figure*}

For our images in each of the five filters, we compare 2 million models to the data using this MCMC method.
The probability distribution for each of the seven
parameters is given by the histogram of the 1-D array (a column of
length 2 million) corresponding 
to the desired model parameter.
The probability distributions from the first and second
halves of the MCMC run are compared to check that the run has converged
to a stable solution. 
In the left panels of Figure~\ref{fig:radec_dists}, we plot the
probability distributions inferred from each of our images separately.
In the right panel, we show the product of the distributions, namely
the cumulative probability distribution. \textcolor{black}{The best estimate is taken
to be the median of the cumulative distribution, while the 1-$\sigma$
uncertainty is given by the range which encloses 34.1\% (total 68.2\%)
of the probability symmetrically on either side of the median.} 
The offset of the ring from the stellar
position in RA is in good agreement for all filters, except $J$-band,
where it differs by half a NICI pixel (\app10 mas). 
This is because the star was 3--5 pixels offset from the
center of the mask during the $J$-band observations, which likely led to
a systematic error in our centroid estimate. 
The offset of the ring in DEC is in good agreement for all five filters. The cumulative probability distributions for the two 
parameters show that the ring offset is 16.7\pp1.3 mas, while the ring
mean radius is estimated to be 1067\pp2 mas. Thus the ratio of
the pericenter distance to the apocenter distance is 0.969\pp0.004 ($\frac{radius-offset}{radius+offset}$).
\textcolor{black}{The PA of the disk offset (26.3\pp3.1\dg ) is roughly along the PA of the ring (26.47\pp0.04\dg ) and the line connecting the NE
and SW brightness peaks}, such that the brighter ansa (NE) is closer to the star, giving credence
to the idea of pericenter glow \citep{1999ApJ...527..918W}. In
Figure~\ref{fig:other_dists}, we also see that the probability distributions for the inclination and PA
of the ring are all in agreement with previous estimates of the ring
properties (see Table~\ref{tab:prevobs}), though all our new measurements are
at higher precision than earlier estimates.
 
\subsection{Method 3: Modeling of Self-Subtraction}

To investigate the possible systematic change in the ring morphology
induced by self-subtraction of flux due to the ADI processing, 
we also tried a modified MCMC approach. In this
method, each time a model image is constructed, a stack of the model 
image rotated to different PAs is also constructed. \textcolor{black}{These PAs are
chosen to be the same as the PAs of the individual images} in the 
real ADI sequence, as they would appear in the reference PSF used for subtraction. The median of this stack of rotated models is then 
subtracted from the original model to simulate the self-subtraction
undergone by the reduced image. The resulting model including
self-subtraction is used to compute $\chi^2$. The MCMC method is otherwise conducted 
in the same way.

\begin{figure*}[ht]

  \centerline{
    \vbox{
     \hbox {
        \includegraphics[width=6cm]{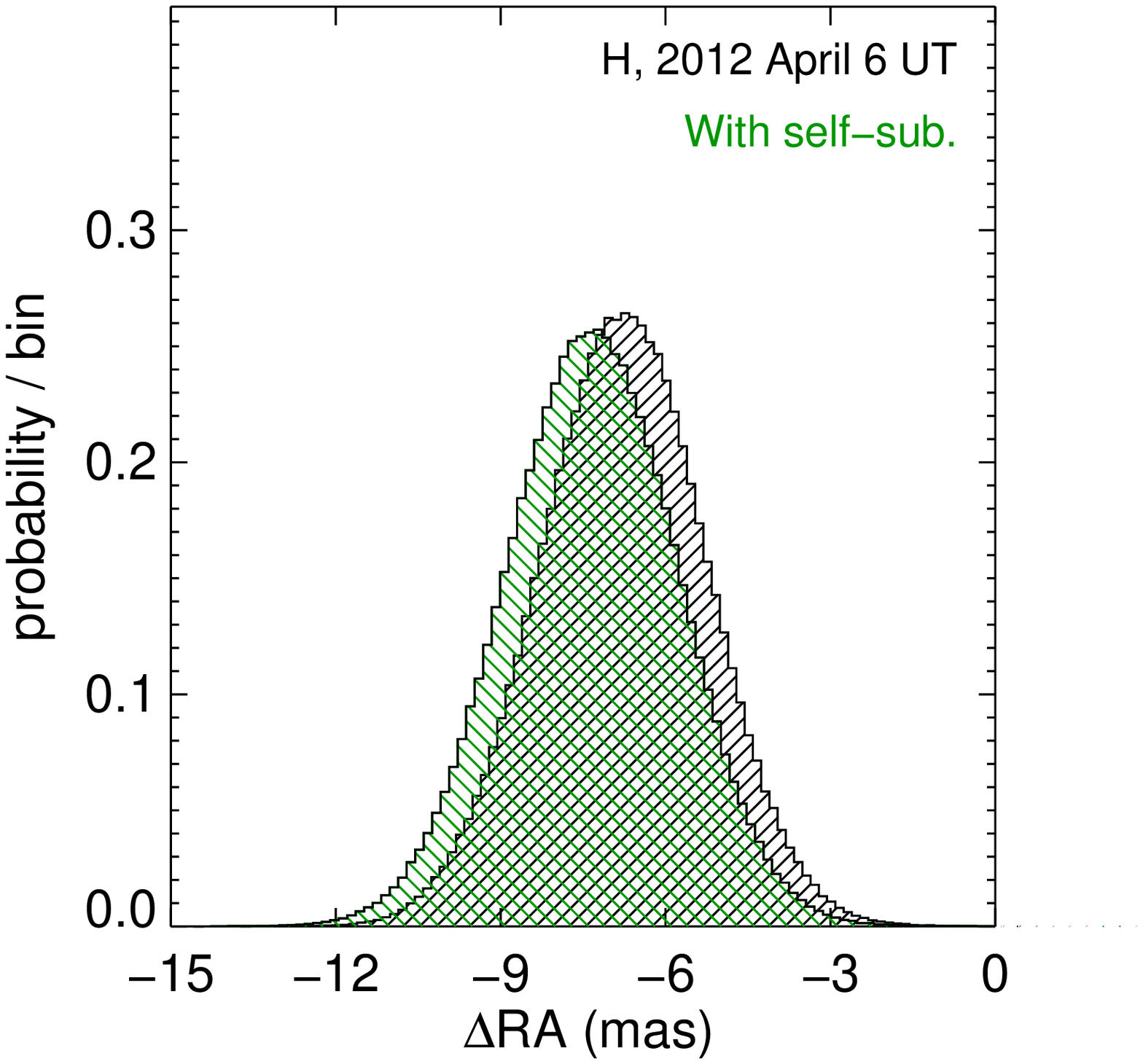}
        \includegraphics[width=6cm]{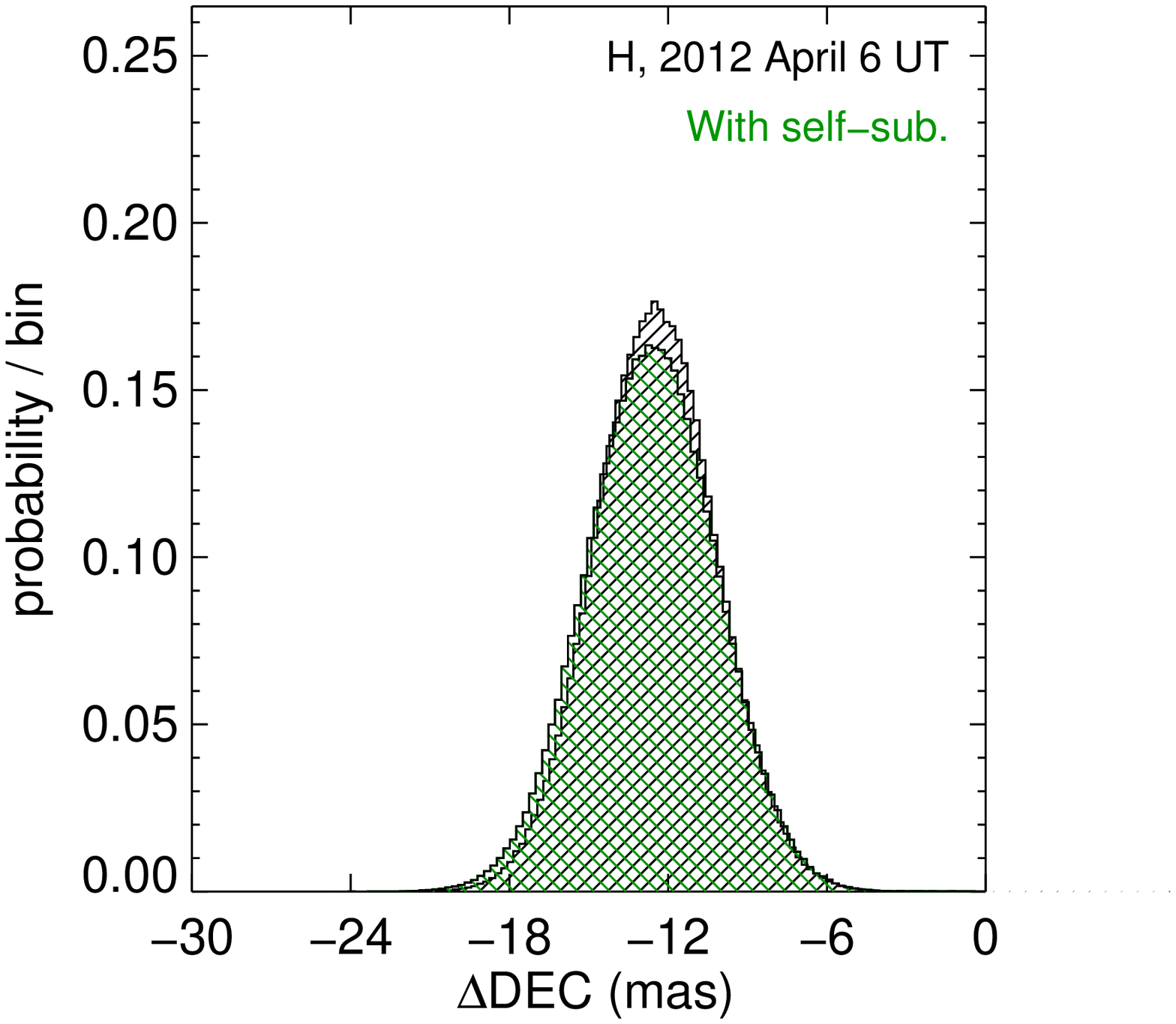}
      }  
     \hbox {
        \includegraphics[width=6cm]{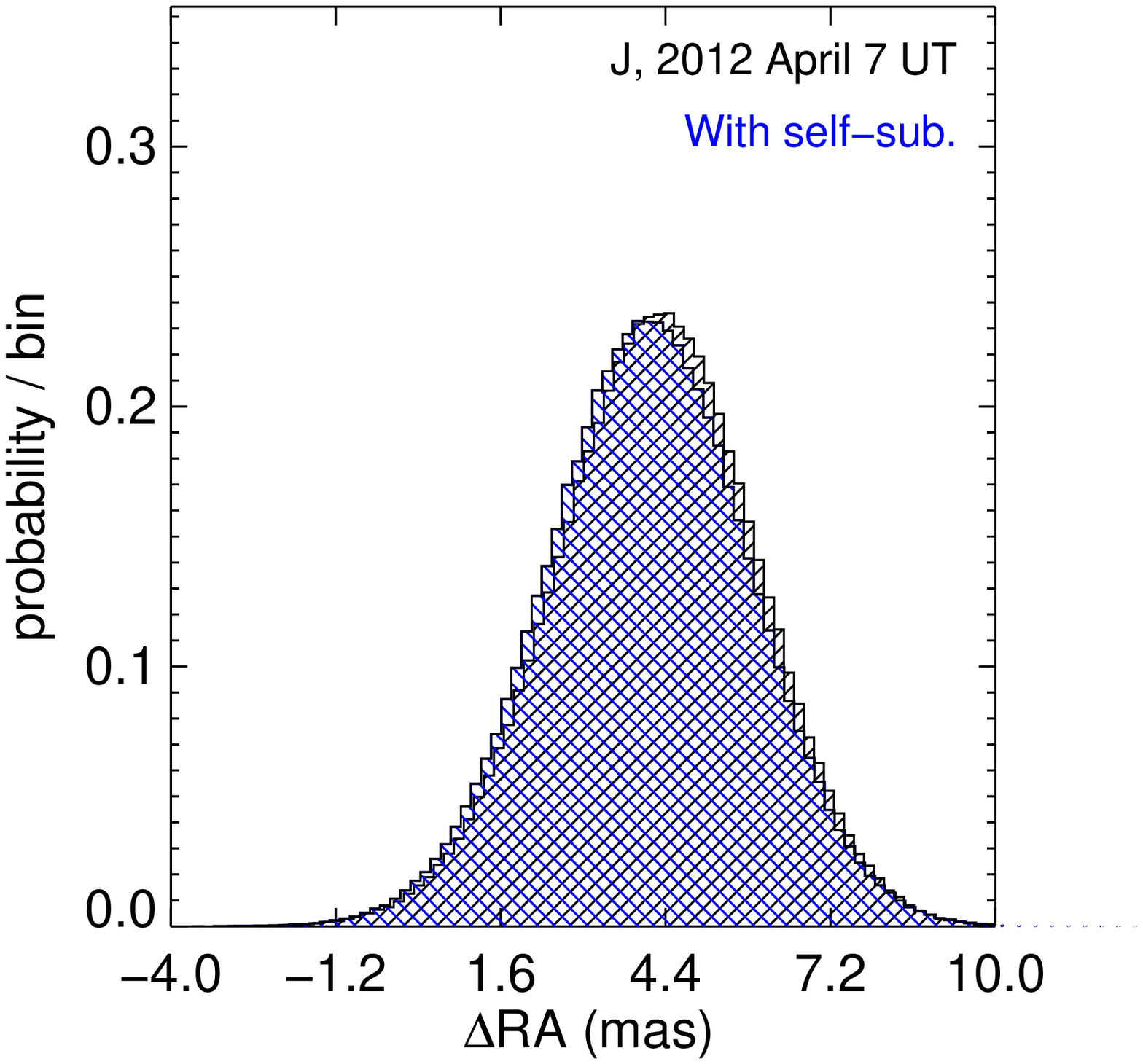}
        \includegraphics[width=6cm]{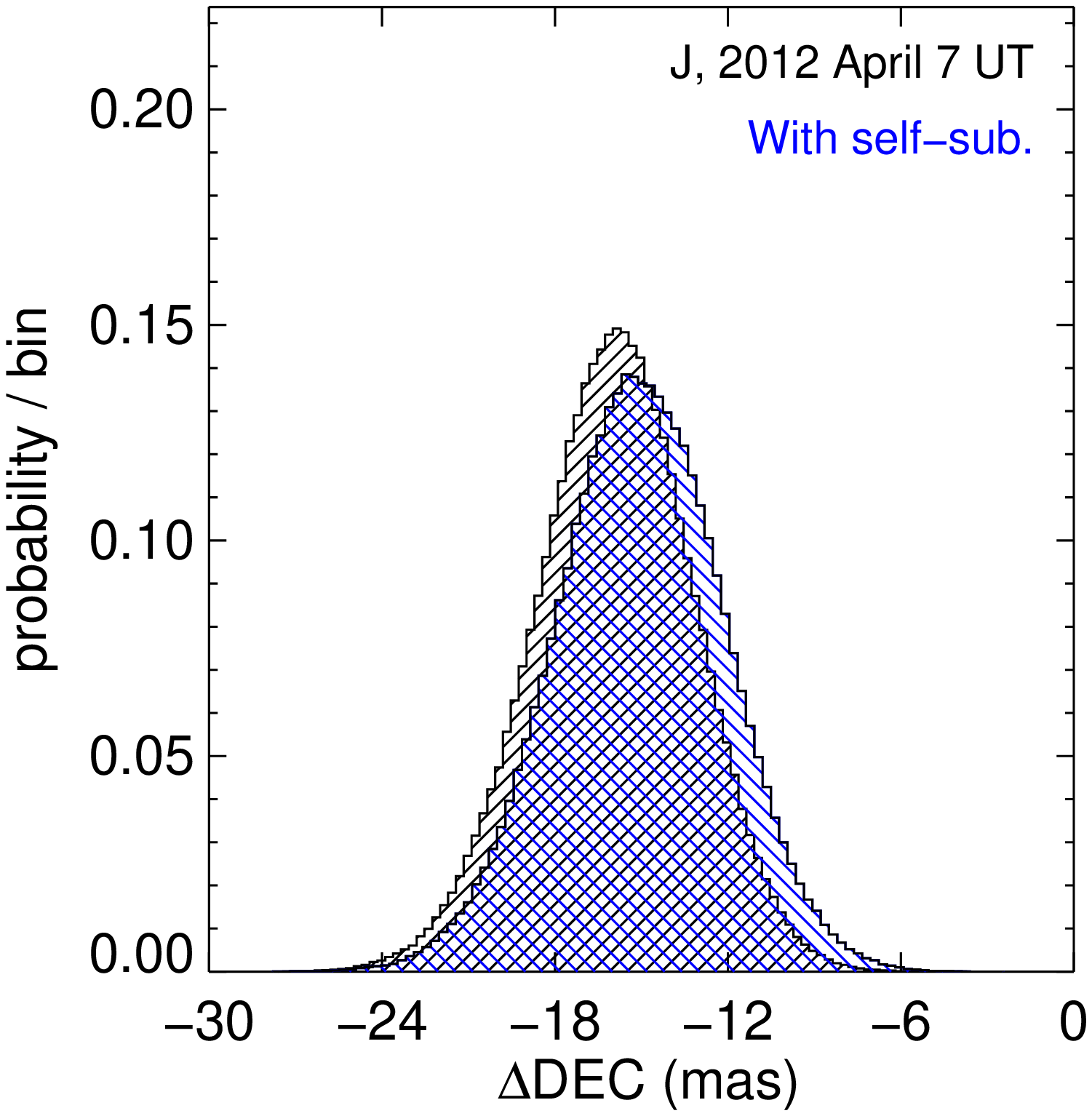}
      }  
      \hbox {
        \includegraphics[width=6cm]{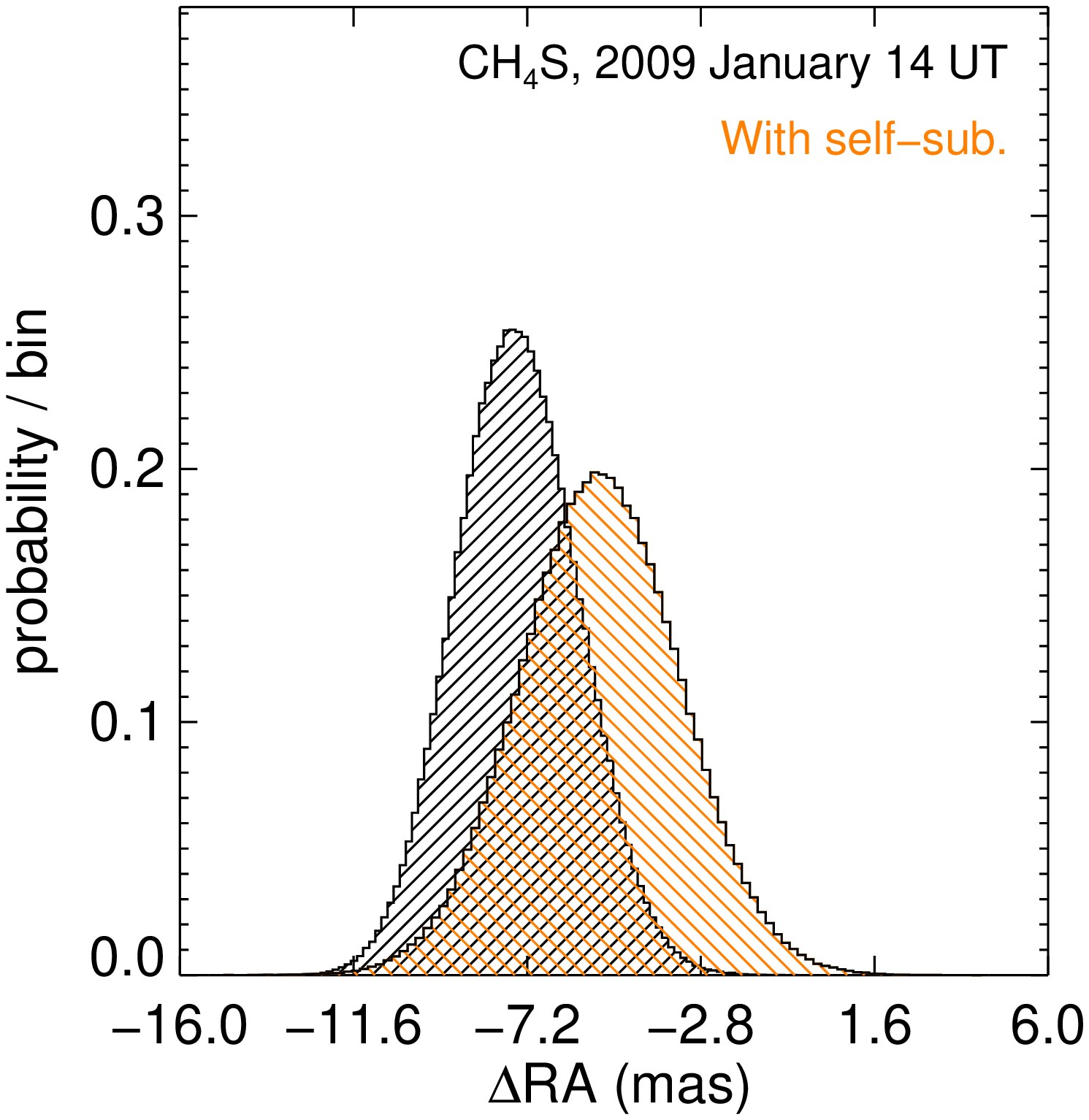}
        \includegraphics[width=6cm]{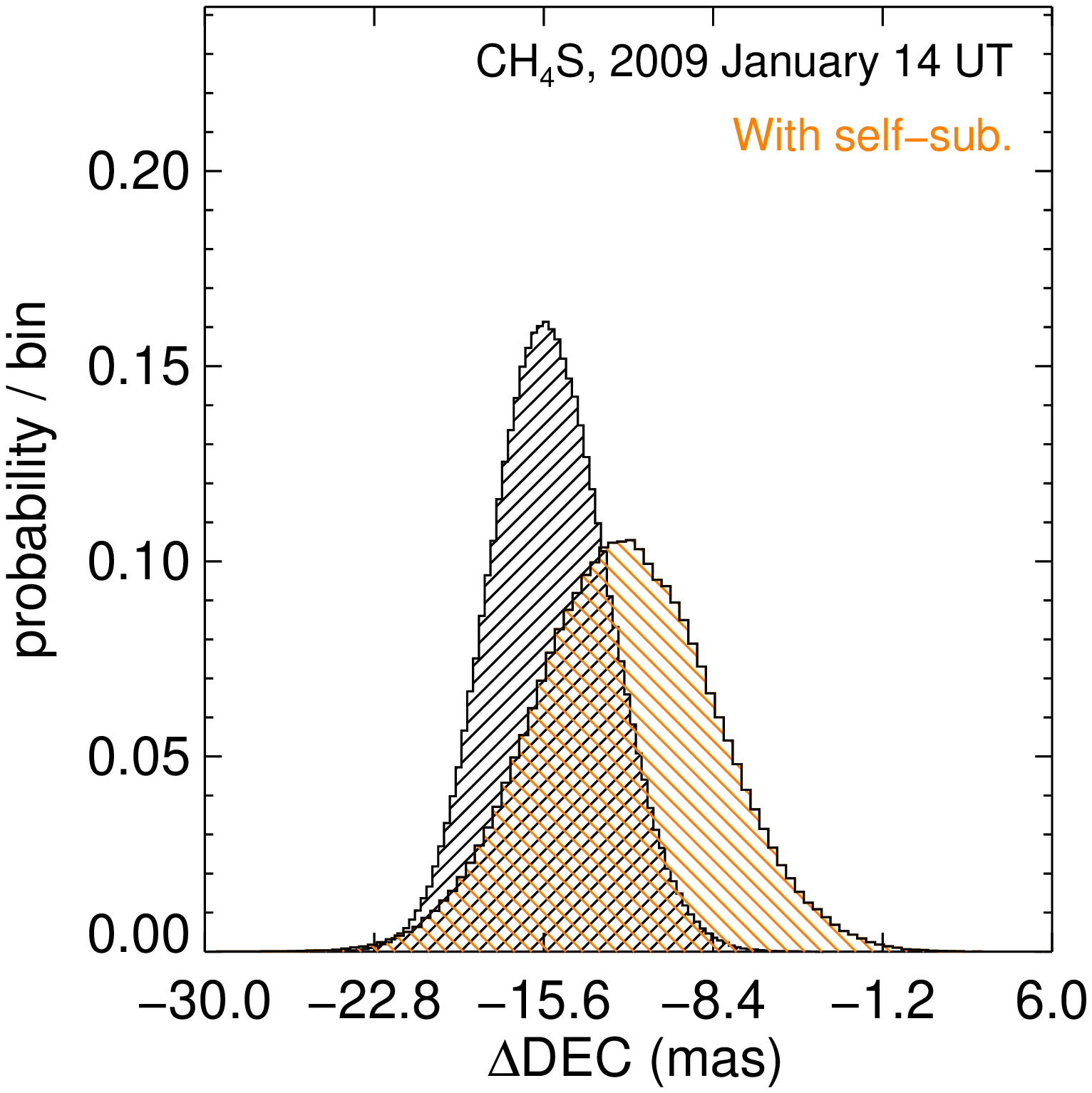}
      }}}  
  \caption{\textcolor{black}{Probability distributions for the $H$-band, 
    $J$-band and $CH_4$ short derived ring
    offsets in RA and DEC for two different MCMC fitting methods. The results from the 
    usual model comparison are shown in black (see section 4.2). The
    results from  the method where the model includes the
    self-subtraction (section 4.3) undergone 
    by the reduced data are shown in green, blue and orange. The offset of the ring from the stellar
    position in RA is in good agreement for all filters, except $J$-band,
    where it differs by half a NICI pixel (\app10 mas). 
    This is because the star was 3--5 pixels offset from the
    center of the mask during the $J$-band observations, which likely led to
    a systematic error in our centroid estimate.} }
  \label{fig:meth3}
\end{figure*}

\textcolor{black}{
The results from method 3 are compared to the usual MCMC approach
(method 2) in Figure~\ref{fig:meth3}. 
\textcolor{black}{The probability distributions from the two methods are in
good agreement for the broad bands  (total sky rotation 73\dg\ for $J$
and 81\dg\ for $H$ and $K_s$) and thus for these bands self-subtraction should not induce 
a large systematic error in the ring offset estimate.} For the $CH_4$-band
data (total sky rotation 24\dg ), the two methods still yield
consistent results but the difference is larger as expected given the
smaller amount of sky rotation.
}

\subsection{Comparison of the Reflectivity of the Dust Ring in $JHK_s$}
\label{sec:jhk}

The observed radiation from the disk in the $JHK_s$ bands is light scattered
from the primary. If we normalize the reduced images by the primary
brightness, then the relative brightnesses of the disk in the
different bands represents the relative reflectivity of the
disk. \textcolor{black}{Should the disk be more reflective in the $J$-band than in the
$H$ or $K_s$-bands, it could indicate the presence of sub-micron
grains \citep[\eg][]{2007ApJ...670..536F,2007ApJ...661..368W}.} 

To better estimate the intensity of the disk, we repeat the reductions
in the $JHK_s$-bands without applying any image filters (step 2 of the
pipeline) to preserve the large-scale emission. \textcolor{black}{We also choose 0.5$''$
wide annuli with radii of $\{$1.8, 1.9, 2.0,
2.1, 2.2$''\}$ (in step 3) when minimizing the residuals from the PSF subtraction} so as to avoid any influence of the disk. 
For the measurements in this section, we estimate uncertainties using the variance of measurements
on the five reductions done with different annuli.  

\begin{figure*}[ht]
  \centerline{
     \hbox {
       \includegraphics[width=8cm]{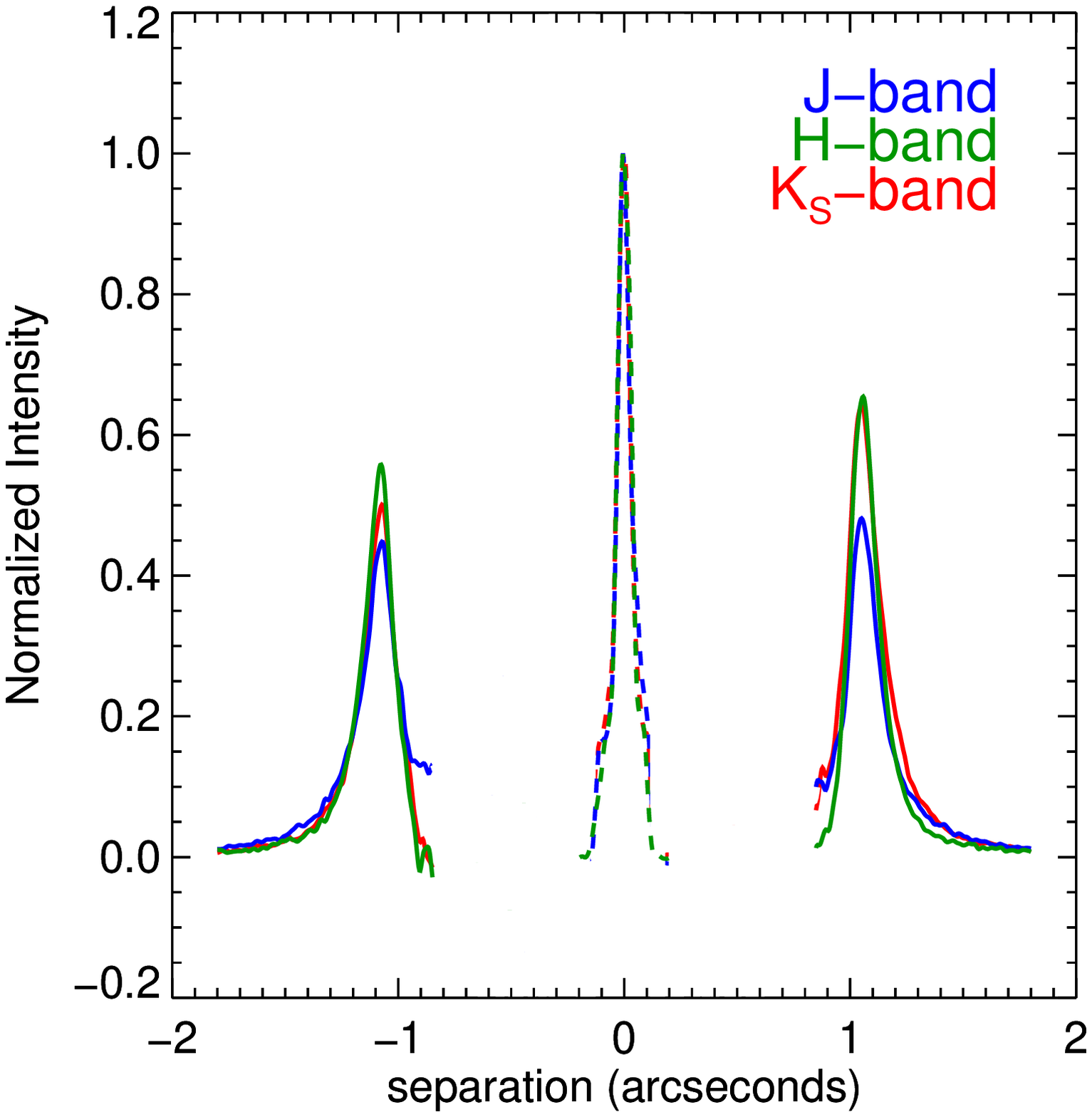}
       \includegraphics[width=8cm]{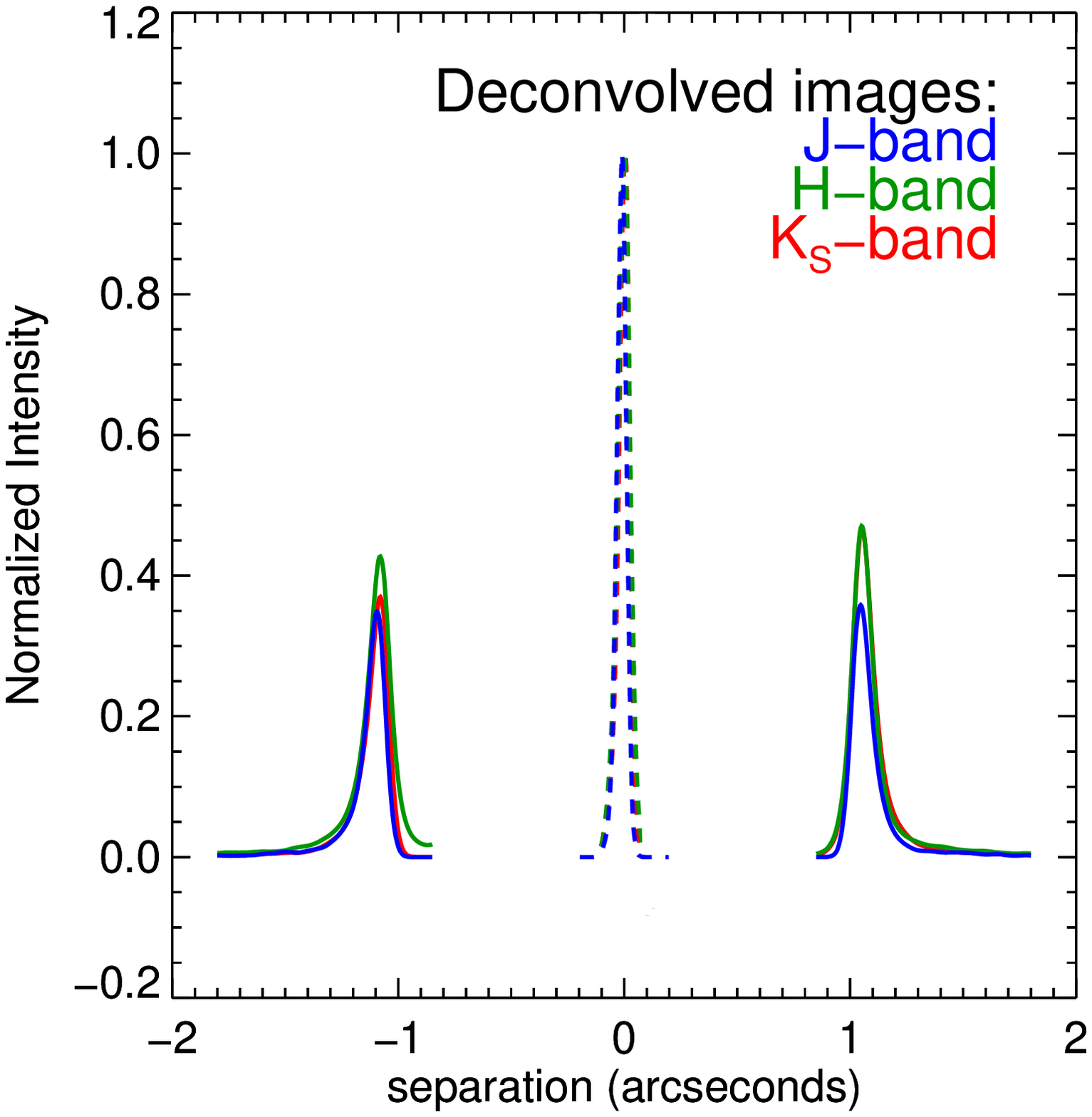}
      }  
    }
    \caption{Left: The intensity profile, normalized by stellar peak, along the PA of the ring in
      the $JHK_s$-bands. The disk has roughly red reflectivity, since
      it is fainter in the $J$-band than the other two bands. Right: The intensity profile of the
      deconvolved versions of the reduced images, made using a
      simultaneously image PSF star in all three bands. The relative disk colors are not
      significantly altered by the different Strehl ratios attained at the three
      wavelengths. In both plots, the dashed lines show the profiles of the stellar
      peak attenuated by a factor of 10000.}
    \label{fig:jhk_prof}
\end{figure*}

To obtain the star-to-ring brightness ratio, we correct for  
the focal plane mask opacity,   6.37 \pp0.12,
 5.94\pp0.05 and 5.70\pp0.09 mag in the $J$, $H$ and $K_s$-bands \citep{2013ApJ...779...80W},
respectively, by multiplying the stellar peak (observed through the
mask) \textcolor{black}{by 10$^{0.4 \times opacity}$}. We then normalize the three images by the corrected stellar
peak and then plot the intensity along the ring's PA as shown in
Figure~\ref{fig:jhk_prof}.
 
\begin{figure*}[ht]
  \centerline{
    \hbox{
      \includegraphics[width=8cm]{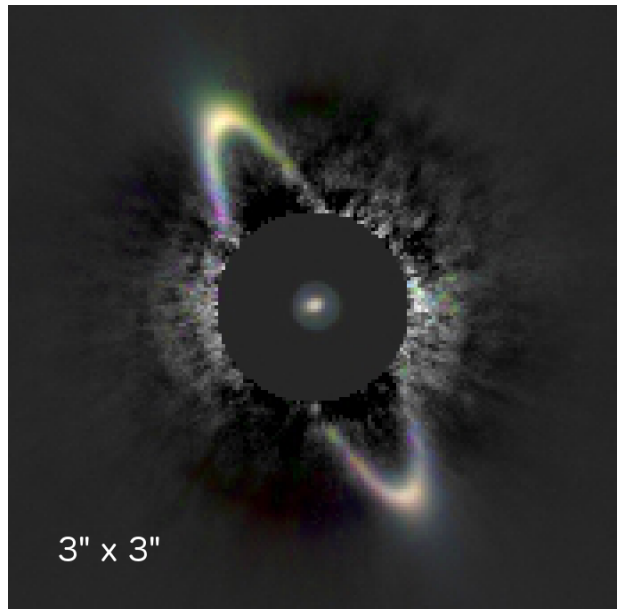}
    }
  }
  \caption{\textcolor{black}{A $JHK_s$ false-color image of the
      HR~4796~A ring, showing its relative reflectivity, as estimated
      in section 4.4. North is up and East is left. The $J$,
    $H$ and $K_s$-bands are colored blue, green and red, respectively.
    The unsaturated star is normalized to one in all the bands and appears
    white in the image. The ring appears yellow because it
    reflects light more efficiently in the $H$ and $K_s$-bands than in
    the $J$-band. The noise-dominated regions with 0.2$''$ to 0.5$''$
    separation from the star are not shown. Regions away from the ring
    are colored white and given the median intensity of the three bands. 
  }}
\label{fig:jhk_falsecol}
\end{figure*}

The ring peak intensity H:J and K:J  ratios on the SW ansa were 1.24\pp0.03
and 1.12\pp0.03, respectively. The H:J and K:J ratios on the NE ansa
were 1.36\pp0.03 and 1.33\pp0.03, respectively. The trend is not clean, but
the disk color seems to be redder than the primary star at least at the peak of the ring.
The relative intensities of the ring are difficult to measure away from the densest 
part of the ring. We find that changing the radius of the fitting annulus has strong effects on
the relative slopes of the ring intensity, thus making the disk colors
highly uncertain away from the peak. \textcolor{black}{Nevertheless, a 
false color image showing the reflectivity of the ring in the three bands is provided in Figure~\ref{fig:jhk_falsecol}}.

\textcolor{black}{
Our ring ansae colors are consistent with
\citet{2008ApJ...673L.191D} in that the H:J and K:J intensity ratios
are between 1.2 and 1.3. However, the \citet{2008ApJ...673L.191D}
NE:SW ansae brightness asymmetries in their F110W,
F160W and F222M bands differ by upto 22\% from those in our $J$, $H$
and $K_S$ bands. We should note that \citet{2008ApJ...673L.191D} calculate total star flux using a standard star and total ring flux using
aperture corrections to estimate the total reflectivity of the
ring. We measure the ring to star contrast by simultaneous unsaturated
imaging of both, and thus comparisons between the two color estimates are difficult.
Detailed spectral modeling efforts \citep[\eg][]{2008ApJ...673L.191D,2008ApJ...686L..95K} to determine the chemical constituents of
the dust would greatly benefit from more accurate color measurements
attainable by the next generation of high-contrast instruments.}     

The different Strehl ratios achieved in the
different bands could in principle contribute to a systematic offset in the star-to-disk 
brightness ratio. \textcolor{black}{Fortunately, we can investigate this possibility using} a known background star, 4.5$''$
away from the primary, which was detected as an unsaturated point source in all three bands.
We deconvolve (16 iterations of Maximum Entropy) the images in each
band with the corresponding point source and compare the
deconvolved images by plotting the intensity profiles along the ring PA, as
shown in Figure~\ref{fig:jhk_prof}.  The deconvolved versions
of the ring retain their star to disk brightness ratios. This indicates that the wavelength-dependent Strehl
ratios do not introduce significant systematics into our measurements.

\subsection{Detection Limits on Planetary Companions}

\textcolor{black}{As we will discuss in the next section, there are dynamical reasons to
believe that there is a planet interior to the dust ring shaping its edges.}
Here we calculate the sensitivity to planets on circular orbits with semi-major axes
\app\ between 50--60~AU in the $H$-band image. To do this, we calculate 
the pixel to pixel RMS within segments of elliptical annulli aligned
with the ring, and calculate the $H$-band contrast limit as\\ $(Stellar\ Peak)/(Mask\ Transmission)/ (10\ \times\ RMS)$, 
where the mask transmission is 0.28\%
\citep{2013ApJ...779...80W}, and the stellar peak is the star's peak flux as measured through the 
coronagraphic mask. We checked by eye that point sources injected
\textcolor{black}{at the annuli location} in the reduced image with peaks at 10\ $\times$\ (local
RMS) are easily discernable. \textcolor{black}{The segments of the elliptical annuli are chosen
to be 30 degrees wide in PA and 10~AU in radial extension, so 
their area in pixels correspond to about 3~NICI resolution elements.
One resolution element is 3 pixels in diameter.} 
We converted the $H$-band contrasts into sensitivity limits in terms of companion
masses, using both COND and DUSTY models
\citep{2000ApJ...542..464C,2003A&A...402..701B}. This is because the
effective temperatures of the planets of the relevant age and intrinsic
brightnesses bracket 1400~K, a temperature below which the dust grains leave the
photosphere.  
 
\begin{figure*}[ht]
  \centerline{
     \hbox {
       \includegraphics[width=8cm]{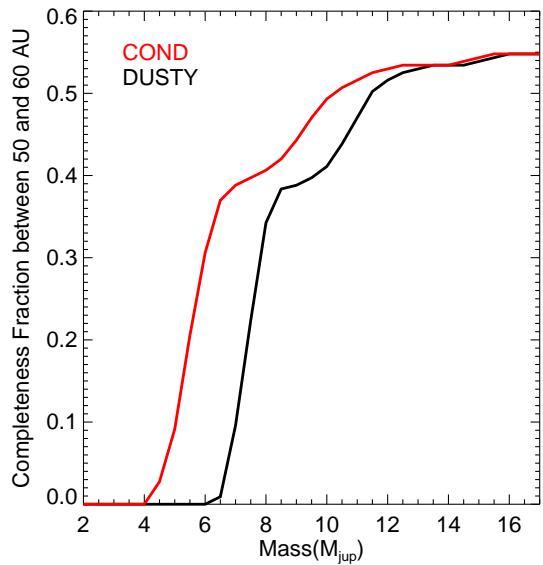}
      }  
    }
    \caption{The fraction of planets, as a function of mass, with
      orbital radii between 50--60~AU (eccentricity $=$ 0) that would be detectable in our
      $H$-band image are shown above, assuming COND (red line) and DUSTY
      (black line) models. The completeness does not reach 100\% as
      some parts of the orbits are under the coronagraphic mask. Neptune-mass planets are massive enough to
      shape the \prim\ ring, but they are beyond our detection limits.}
    \label{fig:complete}
\end{figure*}

We then calculate for a range of planet masses the fraction of the 
orbits where planets would be bright enough to be detected in our $H$-band image. This is then
the completeness as a function of companion mass (Figure~\ref{fig:complete}).  
We only reach significant completeness above 8~\mjup , while only a
Neptune-mass planet is required to carve the edges of the \prim\ ring.

\subsection{Ring Profile in De-projected images}
\textcolor{black}{In \citet{2005ApJ...618..385W}, the authors performed simultaneous modeling of the 
spectral energy distribution, thermal images at 12.5, 20.8 and
24.5~\mic\ and a 1.1~\mic\ scattered light image. 
They found strong evidence for both a tenuous wide and a dense narrow component to 
the \prim\ dust ring.} High precision estimates were made for the
inner and outer edges of the narrow component, 71.7 and 86.9~AU,
with roughly 1\% uncertainty for both. These distances
would match the 2:1 and 3:2 resonances of a hypothetical planet at
54.7~AU. Our NICI images have less fidelity than the {\it Hubble Space
  Telescope} (HST) image used by \citet{2005ApJ...618..385W}, since HST 
has a much more stable PSF and the stellar light is easier to remove
from the images. Thus it would not be very useful to model 
the NICI images without characterizing the systematic errors. 
Instead we simply compare the ring profile in the deconvolved and deprojected NICI
images to the estimated ring edges in \citet{2005ApJ...618..385W}.

\begin{figure*}[ht]
  \centerline{
     \hbox {
       \includegraphics[width=8cm]{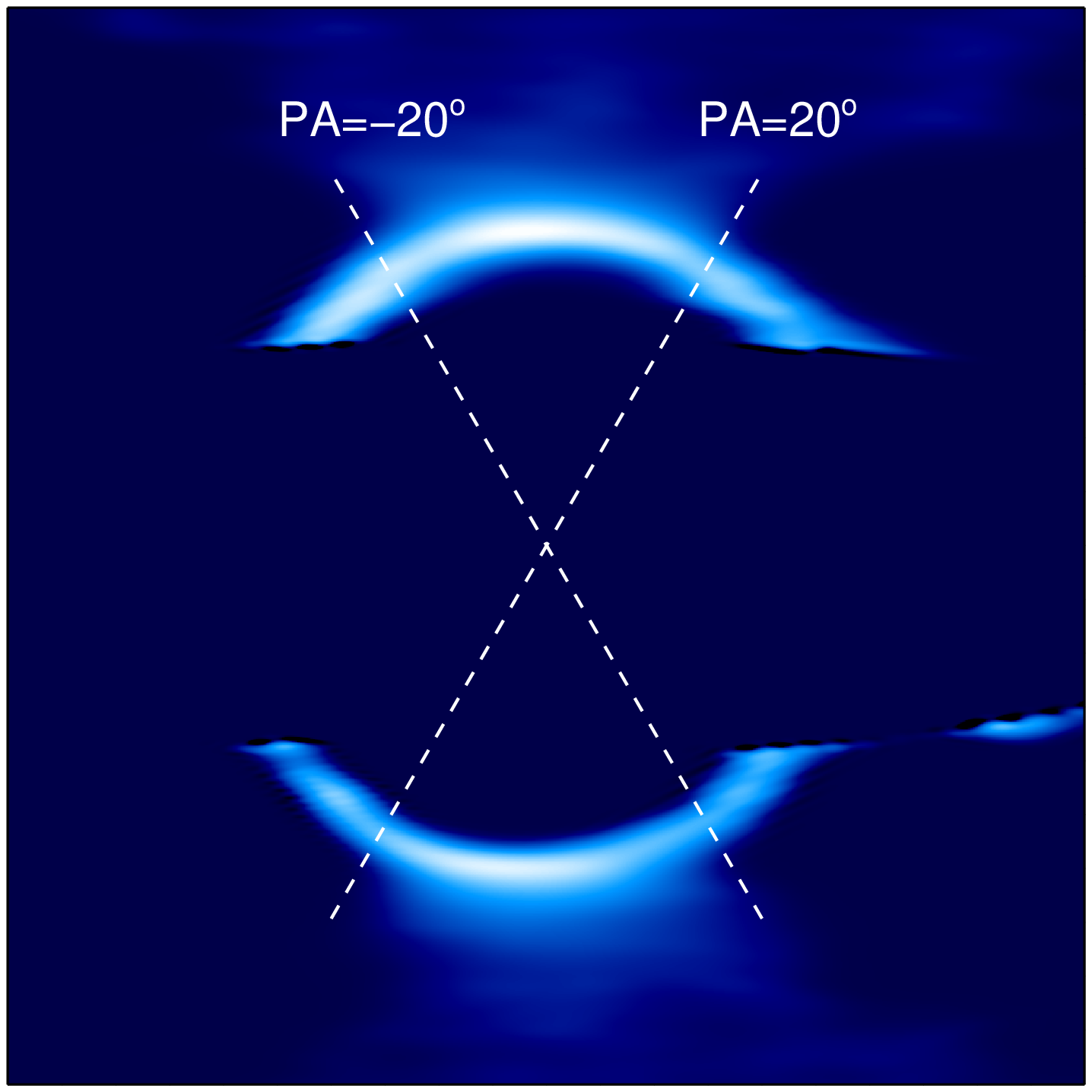}
       \includegraphics[width=8cm]{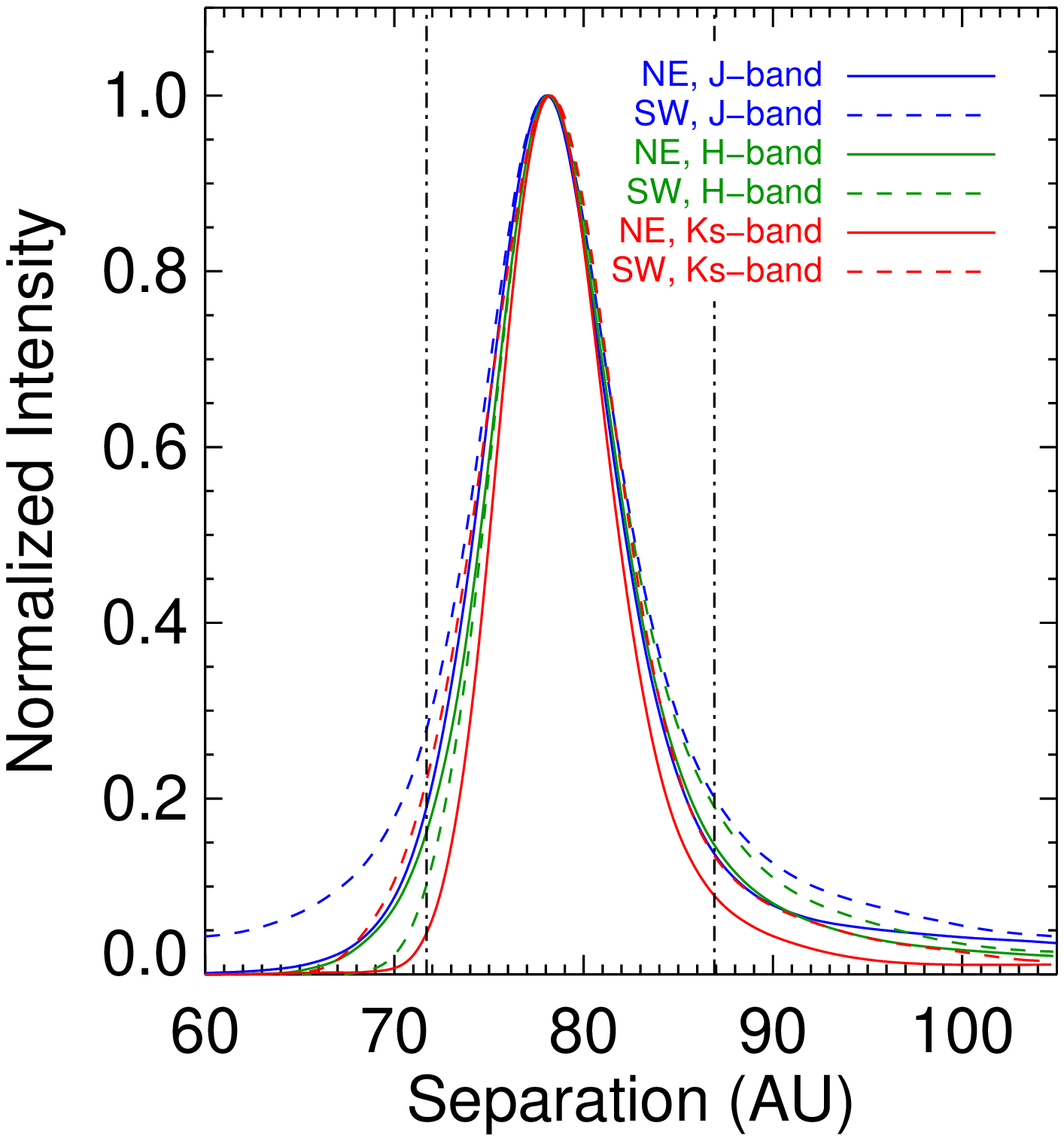}
      }  
    }
    \caption{Left: A de-projection of the deconvolved $H$-band reduced
      image, rotated to make the NE ansa point upwards. The inner,  low signal-to-noise
      region of the ring is not shown here. Only the PA range within
      the two dashed lines are used to created the profiles to the
      right. Right: The $JHKs$-band, normalized profiles of the deconvolved and
      de-projected ring, towards the NE and SW ansae, compared to the inner and outer edge estimates
      for the narrow-ring component in \citet{2005ApJ...618..385W}.}
    \label{fig:deproj}
\end{figure*}

We take our deconvolved images (section \ref{sec:jhk}), and using the known
inclination (76\dg ) and PA (26\dg ), produce the face-on
appearance of the ring. 
This de-projected image in shown in Figure~\ref{fig:deproj} (left
panel), rotated to make the NE ansa point upwards. We plot the mean intensity of
the ring over the PA range $-$20\dg\ to $+$20\dg (in the rotated image) as a function of
separation from the ring center (Figure~\ref{fig:deproj}, right
panel) and compare to the ring edge estimates in \citet{2005ApJ...618..385W}. We calculate that more than 80\% of the ring flux falls within 71.7 and 86.9~AU.
Thus the NICI images are also consistent with the densest part of the
ring being contained within \textcolor{black}{the 2:1 and 3:2 resonances of the
hypothetical planet at 54.7~AU.}

\section{DISCUSSION}

\textcolor{black}{What could be responsible for the brightness
asymmetry we observe in the \prim\ dust ring?} 
Since the ring-to-star distance for the NE and SW ansae have a ratio of
0.97 (NE/SW), in scattered light the SW should be only 0.94 (=0.97$^2$)
times as bright as the NE ansa. 
However, if the dust grains are on elliptical orbits, 
the grain density is also higher by 3\% at the SW ansa, as the
velocity is lower by that much at the apocenter 
($\frac {apocenter~velocity}{pericenter~velocity} \sim ~\frac {pericenter~distance}{apocenter~distance}$ 
according to Kepler's laws). So we are left with 
a naive expectation of  0.97 (0.94/0.97) brightness asymmetry between the ring
ansae. On the other hand, it is expected that higher velocities at the NE ansa will also result in
a higher collision rate \citep{2011AA...526A..34M}.
The brightness asymmetry measurements in scattered light images made so far range from 0.78 to
0.93 (mean= 0.88 from Table~\ref{tab:bright_asymm};
note that the Debes et al.\ 2008\nocite{2008ApJ...673L.191D} 
values are not included due to their large variations at similar wavelengths). 
Thus higher collision rates at the pericenter due to higher velocities
will have to account for a further $\sim$9\% increase in asymmetry to reach the 
mean asymmetry of 0.88 recorded so far.

\textcolor{black}{
How could such a narrow asymmetric dust ring have originated? 
A complete dynamical model should specify the location of the
generators of the dust, the planetesimals, and explain why they are
in a narrow asymmetric ring. The narrowness is important because planetesimal rings are
expected to widen due to velocity dispersion \citep{2007MNRAS.377.1287Q}.
We find a possible solution in \citet{2003ApJ...598.1321W}, who developed a planet migration 
model that is tunable to create different brightness asymmetries in
exterior dust rings by mean-motion resonance trapping. The migration is necessary
to capture a significant number of planetesimals into resonances. Two of the key
parameters in the model are the planet to primary mass ratio and the migration
rate. To match the asymmetry in a submillimeter image of Vega (which has
since been refuted; see Hughes et al.\ 2012\nocite{2012ApJ...750...82H}), \citet{2003ApJ...598.1321W} found that a 
Neptune-mass planet which migrated at a rate of 0.45~AU/Myr from 40 to
65~AU worked well although a range of scenarios were equally
viable.  \prim\ has a similar mass with a similar sized ring, but it is
much younger. From the Wyatt et al.\ simulations we note that only trapping into the 2:1 resonance 
can create the desired asymmetry, since all other resonances produce
clumps of equal brightness. Since the eccentricity of the \prim\ ring is only
0.032, the allowed migration is only 0.1~AU (see their equation 22).
Enough material needs to be present during capture and then
enough time has to elapse for the primodrial to be disperse \citep[2--4~Myr;][]{2010ApJ...724..835W}. Thus
the capture should occur early in the life of the system \citep[8--10~Myr;][]{1995ApJ...454..910S}.  
Moderate migration rates would be consistent with this scenario, say roughly
0.1~AU in 0.1--1~Myr, or 0.1--1~AU/Myr.
Unfortunately, no other constrains can be obtained  by considering this model 
since slow migration rates are consistent with both small and large
planet masses. According to their Figure~4, even a planet as small as
3~\mearth\ would efficiently trap planetesimals into the 2:1
resonance. In any case, a specific simulation should be carried out for
\prim\ to be check if the properties of the ring can be reproduced by
this model.  
}


The outer edge of the disk may be naturally sharp due to several
reasons, although these scenarios have not yet been adequately explored
by dynamical modeling: (1) only a narrow annular region is preserved by
mean-motion resonance with a planet, with the rest of the disk 
dispersed;  (2) only a narrow annular region is dynamically excited by
the mean-motion resonance with the planet; or (3) very recent
planetesimal break-up led to a asymmetric ring-like structure, which has not
diffused yet \citep{2014arXiv1403.1888J}.

An interesting comparison can be made with the connection between the Kuiper Belt and
Neptune in the Solar System. The inner and outer edges
of the classical Kuiper belt coincide with Neptune's 2:1 and 3:2 
resonances, respectively. This phenomenon is poorly understood but
very actively studied \citep{2008Icar..196..258L,2012ApJ...750...43D}. A possible explanation is that Kuiper objects 
were captured in resonance with Neptune as the planet migrated
outward and such objects survived in larger numbers when the Kuiper 
belt was later cleared out by chaotic events. \textcolor{black}{For \prim ,\citet{2005ApJ...618..385W} presented a single dust
disk model consistent with NIR and MIR images, and photometry
from MIR to millimeter wavelengths. In the model, the scattered light emission was dominated by 50~\mic\ grains 
confined to a dense narrow ring, while 7~\mic\ grains formed a tenuous
wide component due to blow-out by radiation pressure and were more
prominent in the MIR.} Using the latest 
distance to \prim\ \citep[72.78~pc;][]{2007A&A...474..653V}, the \citet{2005ApJ...618..385W} estimates for 
the inner and outer edges of the narrow ring are updated to
71.7 and 86.9~AU respectively. These distances correspond to the 
3:2 and 2:1 resonances respectively of a planet at 54.7~AU to within
0.1\%. \textcolor{black}{
However, the precision of the edge radii estimates themselves are only \app1\%.
What is the probability that the edges of the ring would
correspond to these resonances by chance?
For such an orbital solution to exist, the ratio of the inner to outer edge
radius would have to be
0.825~(=$R_{3:2}/R_{2:1}=(P_{3:2}/P_{2:1})^{\frac{2}{3}}$=0.75$^{\frac{2}{3}}$)\pp0.008
(to agree to within 1\%).
Here $R_{2:1}$ and $P_{2:1}$ are the radius and period of the 2:1
resonance of a proposed planet, while $R_{3:2}$ and $P_{3:2}$ correspond
to the 3:2 resonance.
Since primordial disks typically have radii of 100--200~AU
\citep{2009ApJ...701..260I}, the a priori expectation is that the outer
radius should not be much more than twice the inner radius.}
\textcolor{black}{Let us assume that all values of the outer radius ranging from twice
the inner radius to exactly the inner radius have equal probability.
Then, the probability that the inner and outer radius are by accident within 1\%
of the resonance solution we find is also roughly 1\%. 
Thus the existence of a resonant orbital solution may indeed be interesting.} 

We also investigated if the same scenario can provide a viable
explanation for the Fomalhaut
debris disk, the other clearly resolved circumstellar ring system. 
For Fomalhaut, the highest resolution images are from \citet[][ALMA; 870~\mic ]{2012ApJ...750L..21B}
, \citet[][HST; $\lambda$=0.20--1.03~\mic]{2013ApJ...775...56K} and 
\citet[][Herschel; 70 and 160~\mic]{2012A&A...540A.125A}. \citet{2012ApJ...750L..21B} 
made specific measurements of the ring inner and outer edges 
and should be considered the highest precision. The 2$\sigma$ range 
of a Gaussian profile fit to the ring in the calibrated ALMA image was 128~AU
to 155.5~AU (see their Figure 2; right panel). For a planet with orbital radius 97.7~AU, the 3:2 and 
2:1 mean-motion resonances would be at 128~AU and 155.06~AU.
The ALMA image used was both de-projected and primary-beam-corrected.
\citet{2013ApJ...775...56K} does not specifically provide the semi-major axis of
the inner and outer edges of the ring. Thus, we estimated the edge distances 
from their Figure 11, an intensity map along the long axis of the
ring, starting from the stellar position and binned by 20 pixels
along the short axis. We estimated the edge locations where there are clear drops in intensity.
The semi-major axis distances, which were calculated by 
dividing the measured values by (1+eccentricity), were 130 and 156.4~AU.
These estimates also agree with resonant orbital solution to within 1\%.
Lastly, the \citet{2012A&A...540A.125A} estimates, which are from a lower
resolution (5.5'') image, for the semi-major axis distances were 133
and 153~AU. Given that their estimates for the mean semi-major axis of the
elliptical ring has an uncertainty of 1--3~AU, their measurements are
also consistent with the resonant orbit scenario. 

\textcolor{black}{
A more general picture for the properties of planet-sculpted eccentric rings is that
a planet interior to the ring clears a gap delineated by a region of
overlapping mean motion resonances \citep{1980AJ.....85.1122W} 
and that the orbits are collisionally relaxed \citep{2006MNRAS.372L..14Q}.
These constraints yield a unique eccentricity for the perturbing planet and a unique
relationship between the planet's semimajor axis and mass. Refinements on this picture
\citep{2009ApJ...693..734C,2013arXiv1311.1207R} have additionally used the width of the ring to
constrain the planet's mass; more massive planets tend to stir the planetesimals and dust,
producing wider rings. According to the \citet{2013arXiv1311.1207R} 
instructions for calculating a normalized FWHM for the dust ring, our $H$-band images
yield a FWHM of 0.102. As we can see from Figure~\ref{fig:deproj}, the ring
widths in the other bands are very similar to that in the
$H$-band. Our FWHM is smaller than what \citet{2013arXiv1311.1207R} calculated
(0.18) from previous publications on the \prim\ ring, probably because
of better Strehl ratio in the NICI images.
According to equation 5 in \citet{2013arXiv1311.1207R}, the planet would have to have a mass less than 2.4~\mjup\ so as not to make the ring wider.
Its eccentricity would be the same as that of the ring, 0.032. The maximum semi-major axis allowed would be 51.4~AU (using equation 2 in \nocite{2013arXiv1311.1207R}Rodigas et al.\ 2013). But if we
insist that the planet is at 54.7~AU, as in our earlier scenario, then
the mass of the planet has to be less than 1.6~\mjup .}

\textcolor{black}{
There are several alternatives to this one-planet, resonance-overlap model.
Using SMACK, a modeling tool combining dynamical perturbations from
a planet with collisional evolution, \citet{2013ApJ...777..144N} found that differential precession
combined with collisions can play an important role in sculpting narrow rings.
Or conceivably a small, stable amount of undetected circumstellar gas within the ring
could be sculpting the ring via the photoelectric instability described by \citet{2013Natur.499..184L}.
These models require deeper theoretical investigation before comparing their
predictions directly with our data. Also, multiple planets might be involved in sculpting the ring, in which case
there may be no unique predicted configuration for the planetary system.
}

%
%



\section{CONCLUSIONS}

We have determined that the \prim\ ring is offset from the star by
16.7\pp1.3~mas based on unsaturated images in five NIR bands. These images
unambiguously show the offset and confirm earlier lower-precision
measurements \citep{2009AJ....137...53S,2011ApJ...743L...6T}.

\textcolor{black}{The densest part of the ring has a roughly red color 
across the $JHK_s$ filters, indicating 1--5~\mic\ grains
\citep{2008ApJ...673L.191D,2008ApJ...686L..95K} but not 50~\mic\ grains
as in the estimates of \citet{2005ApJ...618..385W}.} Away from the peak density 
we cannot make high precision measurements of the relative
reflectivity of the ring, because of influence of the data reduction 
process on the wings of the ring. 

We show that the brightness asymmetry of the ring in the NIR cannot be
explained by the pericenter-glow effect \citep{1999ApJ...527..918W} alone.
Higher collision rates at the pericenter or some other phenomenon will have to account for 9\% 
additional asymmetry over that provided by the pericenter glow (3\%).

We discuss a possible explanation for the debris disk ring widths, which
has not garnered much attention thus far. Just like Neptune's 3:2 
and 2:1 mean-motion resonances delineate the inner and outer edges 
of the classical Kuiper belt, we find that the radial extent of the
\prim\ and Fomalhaut rings could correspond to 3:2 and 2:1 mean-motion
resonances of hypothetical planets at 54.7~AU and 97.7~AU in the two systems, 
respectively.  For \prim\ we are only sensitive to planets with masses
above several Jupiters. \textcolor{black}{However, a planet at 54.7~AU would have to be
less massive than 1.6~\mjup\ so as not to widen the ring too much by stirring.} 

\begin{acknowledgements}  
This work was supported in part by NSF grants AST-0713881 and
AST-0709484. 
Our research has employed the 2MASS data products; NASA's Astrophysical
Data System; the SIMBAD database operated at CDS, Strasbourg, France.
\end{acknowledgements} 

{\it Facilities:} Gemini-South (NICI), IRTF (SpeX).

\bibliographystyle{aa}
\bibliography{zrefs}

\begin{thebibliography}{58}
\expandafter\ifx\csname natexlab\endcsname\relax\def\natexlab#1{#1}\fi

\bibitem[{{Acke} {et~al.}(2012){Acke}, {Min}, {Dominik}, {Vandenbussche},
  {Sibthorpe}, {Waelkens}, {Olofsson}, {Degroote}, {Smolders}, {Pantin},
  {Barlow}, {Blommaert}, {Brandeker}, {De Meester}, {Dent}, {Exter}, {Di
  Francesco}, {Fridlund}, {Gear}, {Glauser}, {Greaves}, {Harvey}, {Henning},
  {Hogerheijde}, {Holland}, {Huygen}, {Ivison}, {Jean}, {Liseau}, {Naylor},
  {Pilbratt}, {Polehampton}, {Regibo}, {Royer}, {Sicilia-Aguilar}, \&
  {Swinyard}}]{2012A&A...540A.125A}
{Acke}, B., {Min}, M., {Dominik}, C., {et~al.} 2012, \aap, 540, A125

\bibitem[{{Backman} \& {Paresce}(1993)}]{1993prpl.conf.1253B}
{Backman}, D.~E. \& {Paresce}, F. 1993, in Protostars and Planets III, ed.
  E.~H. {Levy} \& J.~I. {Lunine}, 1253--1304

\bibitem[{{Baraffe} {et~al.}(2003){Baraffe}, {Chabrier}, {Barman}, {Allard}, \&
  {Hauschildt}}]{2003A&A...402..701B}
{Baraffe}, I., {Chabrier}, G., {Barman}, T.~S., {Allard}, F., \& {Hauschildt},
  P.~H. 2003, \aap, 402, 701

\bibitem[{{Barrado Y Navascu{\'e}s}(2006)}]{2006A&A...459..511B}
{Barrado Y Navascu{\'e}s}, D. 2006, \aap, 459, 511

\bibitem[{{Biller} {et~al.}(2013){Biller}, {Liu}, {Wahhaj}, {Nielsen},
  {Hayward}, {Males}, {Skemer}, {Close}, {Chun}, {Ftaclas}, {Clarke}, {Thatte},
  {Shkolnik}, {Reid}, {Hartung}, {Boss}, {Lin}, {Alencar}, {de Gouveia Dal
  Pino}, {Gregorio-Hetem}, \& {Toomey}}]{2013ApJ...777..160B}
{Biller}, B.~A., {Liu}, M.~C., {Wahhaj}, Z., {et~al.} 2013, \apj, 777, 160

\bibitem[{{Boley} {et~al.}(2012){Boley}, {Payne}, {Corder}, {Dent}, {Ford}, \&
  {Shabram}}]{2012ApJ...750L..21B}
{Boley}, A.~C., {Payne}, M.~J., {Corder}, S., {et~al.} 2012, \apjl, 750, L21

\bibitem[{{Chabrier} {et~al.}(2000){Chabrier}, {Baraffe}, {Allard}, \&
  {Hauschildt}}]{2000ApJ...542..464C}
{Chabrier}, G., {Baraffe}, I., {Allard}, F., \& {Hauschildt}, P. 2000, \apj,
  542, 464

\bibitem[{{Chiang} {et~al.}(2009){Chiang}, {Kite}, {Kalas}, {Graham}, \&
  {Clampin}}]{2009ApJ...693..734C}
{Chiang}, E., {Kite}, E., {Kalas}, P., {Graham}, J.~R., \& {Clampin}, M. 2009,
  \apj, 693, 734

\bibitem[{{Chun} {et~al.}(2008){Chun}, {Toomey}, {Wahhaj}, {Biller}, {Artigau},
  {Hayward}, {Liu}, {Close}, {Hartung}, {Rigaut}, \&
  {Ftaclas}}]{2008SPIE.7015E..49C}
{Chun}, M., {Toomey}, D., {Wahhaj}, Z., {et~al.} 2008, in Society of
  Photo-Optical Instrumentation Engineers (SPIE) Conference Series, Vol. 7015,
  Society of Photo-Optical Instrumentation Engineers (SPIE) Conference Series

\bibitem[{{Dawson} \& {Murray-Clay}(2012)}]{2012ApJ...750...43D}
{Dawson}, R.~I. \& {Murray-Clay}, R. 2012, \apj, 750, 43

\bibitem[{{Debes} {et~al.}(2008){Debes}, {Weinberger}, \&
  {Schneider}}]{2008ApJ...673L.191D}
{Debes}, J.~H., {Weinberger}, A.~J., \& {Schneider}, G. 2008, \apjl, 673, L191

\bibitem[{{Fitzgerald} {et~al.}(2007){Fitzgerald}, {Kalas}, {Duch{\^e}ne},
  {Pinte}, \& {Graham}}]{2007ApJ...670..536F}
{Fitzgerald}, M.~P., {Kalas}, P.~G., {Duch{\^e}ne}, G., {Pinte}, C., \&
  {Graham}, J.~R. 2007, \apj, 670, 536

\bibitem[{Gregory(2005)}]{Gregory05}
Gregory, P.~C. 2005, Bayesian Logical Data Analysis for the Physical Sciences:
  A Comparative Approach with Mathematica Support (book) (New York, NY, USA:
  Cambridge University Press)

\bibitem[{{Heap} {et~al.}(2000){Heap}, {Lindler}, {Lanz}, {Cornett}, {Hubeny},
  {Maran}, \& {Woodgate}}]{2000ApJ...539..435H}
{Heap}, S.~R., {Lindler}, D.~J., {Lanz}, T.~M., {et~al.} 2000, \apj, 539, 435

\bibitem[{{Hughes} {et~al.}(2012){Hughes}, {Wilner}, {Mason}, {Carpenter},
  {Plambeck}, {Chiang}, {Andrews}, {Williams}, {Hales}, {Su}, {Chiang},
  {Dicker}, {Korngut}, \& {Devlin}}]{2012ApJ...750...82H}
{Hughes}, A.~M., {Wilner}, D.~J., {Mason}, B., {et~al.} 2012, \apj, 750, 82

\bibitem[{{Isella} {et~al.}(2009){Isella}, {Carpenter}, \&
  {Sargent}}]{2009ApJ...701..260I}
{Isella}, A., {Carpenter}, J.~M., \& {Sargent}, A.~I. 2009, \apj, 701, 260

\bibitem[{{Jackson} {et~al.}(2014){Jackson}, {Wyatt}, {Bonsor}, \&
  {Veras}}]{2014arXiv1403.1888J}
{Jackson}, A.~P., {Wyatt}, M.~C., {Bonsor}, A., \& {Veras}, D. 2014, ArXiv
  e-prints

\bibitem[{{Jura} {et~al.}(1993){Jura}, {Zuckerman}, {Becklin}, \&
  {Smith}}]{1993ApJ...418L..37J}
{Jura}, M., {Zuckerman}, B., {Becklin}, E.~E., \& {Smith}, R.~C. 1993, \apjl,
  418, L37

\bibitem[{{Kalas} {et~al.}(2013){Kalas}, {Graham}, {Fitzgerald}, \&
  {Clampin}}]{2013ApJ...775...56K}
{Kalas}, P., {Graham}, J.~R., {Fitzgerald}, M.~P., \& {Clampin}, M. 2013, \apj,
  775, 56

\bibitem[{{Koerner} {et~al.}(1998){Koerner}, {Ressler}, {Werner}, \&
  {Backman}}]{1998ApJ...503L..83K}
{Koerner}, D.~W., {Ressler}, M.~E., {Werner}, M.~W., \& {Backman}, D.~E. 1998,
  \apjl, 503, L83

\bibitem[{{K{\"o}hler} {et~al.}(2008){K{\"o}hler}, {Mann}, \&
  {Li}}]{2008ApJ...686L..95K}
{K{\"o}hler}, M., {Mann}, I., \& {Li}, A. 2008, \apjl, 686, L95

\bibitem[{{Lagrange} {et~al.}(2009){Lagrange}, {Gratadour}, {Chauvin}, {Fusco},
  {Ehrenreich}, {Mouillet}, {Rousset}, {Rouan}, {Allard}, {Gendron}, {Charton},
  {Mugnier}, {Rabou}, {Montri}, \& {Lacombe}}]{2009A&A...493L..21L}
{Lagrange}, A., {Gratadour}, D., {Chauvin}, G., {et~al.} 2009, \aap, 493, L21

\bibitem[{{Lagrange} {et~al.}(2012){Lagrange}, {Boccaletti}, {Milli},
  {Chauvin}, \& {Others}}]{2012A&A...542A..40L}
{Lagrange}, A.-M., {Boccaletti}, A., {Milli}, J., {Chauvin}, G., \& {Others}.
  2012, \aap, 542, A40

\bibitem[{{Lagrange} {et~al.}(2010){Lagrange}, {Bonnefoy}, {Chauvin}, {Apai},
  {Ehrenreich}, {Boccaletti}, {Gratadour}, {Rouan}, {Mouillet}, {Lacour}, \&
  {Kasper}}]{2010Sci...329...57L}
{Lagrange}, A.-M., {Bonnefoy}, M., {Chauvin}, G., {et~al.} 2010, Science, 329,
  57

\bibitem[{{Levison} {et~al.}(2008){Levison}, {Morbidelli}, {Van Laerhoven},
  {Gomes}, \& {Tsiganis}}]{2008Icar..196..258L}
{Levison}, H.~F., {Morbidelli}, A., {Van Laerhoven}, C., {Gomes}, R., \&
  {Tsiganis}, K. 2008, \icarus, 196, 258

\bibitem[{{Liu}(2004)}]{2004Sci...305.1442L}
{Liu}, M.~C. 2004, Science, 305, 1442

\bibitem[{{Liu} {et~al.}(2010){Liu}, {Wahhaj}, {Biller}, {Nielsen}, {Chun},
  {Close}, {Ftaclas}, {Hartung}, {Hayward}, {Clarke}, {Reid}, {Shkolnik},
  {Tecza}, {Thatte}, {Alencar}, {Artymowicz}, {Boss}, {Burrows}, {de Gouveia
  Dal Pino}, {Gregorio-Hetem}, {Ida}, {Kuchner}, {Lin}, \&
  {Toomey}}]{2010SPIE.7736E..53L}
{Liu}, M.~C., {Wahhaj}, Z., {Biller}, B.~A., {et~al.} 2010, in Society of
  Photo-Optical Instrumentation Engineers (SPIE) Conference Series, Vol. 7736,
  Society of Photo-Optical Instrumentation Engineers (SPIE) Conference Series

\bibitem[{{Lyra} \& {Kuchner}(2013)}]{2013Natur.499..184L}
{Lyra}, W. \& {Kuchner}, M. 2013, \nat, 499, 184

\bibitem[{{Marois} {et~al.}(2006){Marois}, {Lafreni{\`e}re}, {Doyon},
  {Macintosh}, \& {Nadeau}}]{2006ApJ...641..556M}
{Marois}, C., {Lafreni{\`e}re}, D., {Doyon}, R., {Macintosh}, B., \& {Nadeau},
  D. 2006, \apj, 641, 556

\bibitem[{{Marois} {et~al.}(2010){Marois}, {Zuckerman}, {Konopacky},
  {Macintosh}, \& {Barman}}]{2010Natur.468.1080M}
{Marois}, C., {Zuckerman}, B., {Konopacky}, Q.~M., {Macintosh}, B., \&
  {Barman}, T. 2010, \nat, 468, 1080

\bibitem[{{Metropolis} {et~al.}(1953){Metropolis}, {Rosenbluth}, {Rosenbluth},
  {Teller}, \& {Teller}}]{1953JChPh..21.1087M}
{Metropolis}, N., {Rosenbluth}, A.~W., {Rosenbluth}, M.~N., {Teller}, A.~H., \&
  {Teller}, E. 1953, \jcp, 21, 1087

\bibitem[{{Milli} {et~al.}(2012){Milli}, {Mouillet}, {Lagrange}, {Boccaletti},
  {Mawet}, {Chauvin}, \& {Bonnefoy}}]{2012A&A...545A.111M}
{Milli}, J., {Mouillet}, D., {Lagrange}, A.-M., {et~al.} 2012, \aap, 545, A111

\bibitem[{{Moerchen} {et~al.}(2011){Moerchen}, {Churcher}, {Telesco}, {Wyatt},
  {Fisher}, \& {Packham}}]{2011AA...526A..34M}
{Moerchen}, M.~M., {Churcher}, L.~J., {Telesco}, C.~M., {et~al.} 2011, \aap,
  526, A34

\bibitem[{{Mouillet} {et~al.}(1997){Mouillet}, {Larwood}, {Papaloizou}, \&
  {Lagrange}}]{1997MNRAS.292..896M}
{Mouillet}, D., {Larwood}, J.~D., {Papaloizou}, J.~C.~B., \& {Lagrange}, A.~M.
  1997, \mnras, 292, 896

\bibitem[{{Nesvold} {et~al.}(2013){Nesvold}, {Kuchner}, {Rein}, \&
  {Pan}}]{2013ApJ...777..144N}
{Nesvold}, E.~R., {Kuchner}, M.~J., {Rein}, H., \& {Pan}, M. 2013, \apj, 777,
  144

\bibitem[{{Nielsen} {et~al.}(2013){Nielsen}, {Liu}, {Wahhaj}, {Biller},
  {Hayward}, {Close}, {Males}, {Skemer}, {Chun}, {Ftaclas}, {Alencar},
  {Artymowicz}, {Boss}, {Clarke}, {de Gouveia Dal Pino}, {Gregorio-Hetem},
  {Hartung}, {Ida}, {Kuchner}, {Lin}, {Reid}, {Shkolnik}, {Tecza}, {Thatte}, \&
  {Toomey}}]{2013ApJ...776....4N}
{Nielsen}, E.~L., {Liu}, M.~C., {Wahhaj}, Z., {et~al.} 2013, \apj, 776, 4

\bibitem[{{Press} {et~al.}(1992){Press}, {Teukolsky}, {Vetterling}, \&
  {Flannery}}]{1992nrca.book.....P}
{Press}, W.~H., {Teukolsky}, S.~A., {Vetterling}, W.~T., \& {Flannery}, B.~P.
  1992, {Numerical recipes in C. The art of scientific computing}, ed. {Press,
  W.~H., Teukolsky, S.~A., Vetterling, W.~T., \& Flannery, B.~P. }

\bibitem[{{Quillen}(2007)}]{2007MNRAS.377.1287Q}
{Quillen}, A. 2007, \mnras, 377, 1287

\bibitem[{{Quillen}(2006)}]{2006MNRAS.372L..14Q}
{Quillen}, A.~C. 2006, \mnras, 372, L14

\bibitem[{{Rameau} {et~al.}(2013){Rameau}, {Chauvin}, {Lagrange}, {Meshkat},
  {Boccaletti}, {Quanz}, {Currie}, {Mawet}, {Girard}, {Bonnefoy}, \&
  {Kenworthy}}]{2013ApJ...779L..26R}
{Rameau}, J., {Chauvin}, G., {Lagrange}, A.-M., {et~al.} 2013, \apjl, 779, L26

\bibitem[{{Rodigas} {et~al.}(2013){Rodigas}, {Malhotra}, \&
  {Hinz}}]{2013arXiv1311.1207R}
{Rodigas}, T.~J., {Malhotra}, R., \& {Hinz}, P.~M. 2013, ArXiv e-prints

\bibitem[{{Schneider} {et~al.}(1999){Schneider}, {Smith}, {Becklin}, {Koerner},
  {Meier}, {Hines}, {Lowrance}, {Terrile}, {Thompson}, \&
  {Rieke}}]{1999ApJ...513L.127S}
{Schneider}, G., {Smith}, B.~A., {Becklin}, E.~E., {et~al.} 1999, \apjl, 513,
  L127

\bibitem[{{Schneider} {et~al.}(2009){Schneider}, {Weinberger}, {Becklin},
  {Debes}, \& {Smith}}]{2009AJ....137...53S}
{Schneider}, G., {Weinberger}, A.~J., {Becklin}, E.~E., {Debes}, J.~H., \&
  {Smith}, B.~A. 2009, \aj, 137, 53

\bibitem[{{Smith} \& {Terrile}(1984)}]{1984Sci...226.1421S}
{Smith}, B.~A. \& {Terrile}, R.~J. 1984, Science, 226, 1421

\bibitem[{{Stauffer} {et~al.}(1995){Stauffer}, {Hartmann}, \& {Barrado y
  Navascues}}]{1995ApJ...454..910S}
{Stauffer}, J.~R., {Hartmann}, L.~W., \& {Barrado y Navascues}, D. 1995, \apj,
  454, 910

\bibitem[{{Thalmann} {et~al.}(2011){Thalmann}, {Janson}, {Buenzli}, {Brandt},
  {Wisniewski}, {Moro-Mart{\'{\i}}n}, {Usuda}, {Schneider}, {Carson},
  {McElwain}, {Grady}, {Goto}, {Abe}, {Brandner}, {Dominik}, {Egner}, {Feldt},
  {Fukue}, {Golota}, {Guyon}, {Hashimoto}, {Hayano}, {Hayashi}, {Hayashi},
  {Henning}, {Hodapp}, {Ishii}, {Iye}, {Kandori}, {Knapp}, {Kudo}, {Kusakabe},
  {Kuzuhara}, {Matsuo}, {Miyama}, {Morino}, {Nishimura}, {Pyo}, {Serabyn},
  {Suto}, {Suzuki}, {Takahashi}, {Takami}, {Takato}, {Terada}, {Tomono},
  {Turner}, {Watanabe}, {Yamada}, {Takami}, \& {Tamura}}]{2011ApJ...743L...6T}
{Thalmann}, C., {Janson}, M., {Buenzli}, E., {et~al.} 2011, \apjl, 743, L6

\bibitem[{{Th{\'e}bault} {et~al.}(2010){Th{\'e}bault}, {Marzari}, \&
  {Augereau}}]{2010A&A...524A..13T}
{Th{\'e}bault}, P., {Marzari}, F., \& {Augereau}, J.-C. 2010, \aap, 524, A13

\bibitem[{{van Leeuwen}(2007)}]{2007A&A...474..653V}
{van Leeuwen}, F. 2007, \aap, 474, 653

\bibitem[{{Wahhaj} {et~al.}(2010){Wahhaj}, {Cieza}, {Koerner}, {Stapelfeldt},
  {Padgett}, {Case}, {Keller}, {Mer{\'{\i}}n}, {Evans}, {Harvey}, {Sargent},
  {van Dishoeck}, {Allen}, {Blake}, {Brooke}, {Chapman}, {Mundy}, \&
  {Myers}}]{2010ApJ...724..835W}
{Wahhaj}, Z., {Cieza}, L., {Koerner}, D.~W., {et~al.} 2010, \apj, 724, 835

\bibitem[{{Wahhaj} {et~al.}(2005){Wahhaj}, {Koerner}, {Backman}, {Werner},
  {Serabyn}, {Ressler}, \& {Lis}}]{2005ApJ...618..385W}
{Wahhaj}, Z., {Koerner}, D.~W., {Backman}, D.~E., {et~al.} 2005, \apj, 618, 385

\bibitem[{{Wahhaj} {et~al.}(2007){Wahhaj}, {Koerner}, \&
  {Sargent}}]{2007ApJ...661..368W}
{Wahhaj}, Z., {Koerner}, D.~W., \& {Sargent}, A.~I. 2007, \apj, 661, 368

\bibitem[{{Wahhaj} {et~al.}(2011){Wahhaj}, {Liu}, {Biller}, {Clarke},
  {Nielsen}, {Close}, {Hayward}, {Mamajek}, {Cushing}, {Dupuy}, {Tecza},
  {Thatte}, {Chun}, {Ftaclas}, {Hartung}, {Reid}, {Shkolnik}, {Alencar},
  {Artymowicz}, {Boss}, {de Gouveia Dal Pino}, {Gregorio-Hetem}, {Ida},
  {Kuchner}, {Lin}, \& {Toomey}}]{2011ApJ...729..139W}
{Wahhaj}, Z., {Liu}, M.~C., {Biller}, B.~A., {et~al.} 2011, \apj, 729, 139

\bibitem[{{Wahhaj} {et~al.}(2013{\natexlab{a}}){Wahhaj}, {Liu}, {Biller},
  {Nielsen}, {Close}, {Hayward}, \& {Others}}]{2013ApJ...779...80W}
{Wahhaj}, Z., {Liu}, M.~C., {Biller}, B.~A., {et~al.} 2013{\natexlab{a}}, \apj,
  779, 80

\bibitem[{{Wahhaj} {et~al.}(2013{\natexlab{b}}){Wahhaj}, {Liu}, {Nielsen},
  {Biller}, {Hayward}, {Close}, \& {Others}}]{2013ApJ...773..179W}
{Wahhaj}, Z., {Liu}, M.~C., {Nielsen}, E.~L., {et~al.} 2013{\natexlab{b}},
  \apj, 773, 179

\bibitem[{{Wisdom}(1980)}]{1980AJ.....85.1122W}
{Wisdom}, J. 1980, \aj, 85, 1122

\bibitem[{{Wyatt}(2008)}]{2008ARA&A..46..339W}
{Wyatt}. 2008, ARA\&A, 46, 339

\bibitem[{{Wyatt}(2003)}]{2003ApJ...598.1321W}
{Wyatt}, M.~C. 2003, \apj, 598, 1321

\bibitem[{{Wyatt} {et~al.}(1999){Wyatt}, {Dermott}, {Telesco}, {Fisher},
  {Grogan}, {Holmes}, \& {Pi{\~n}a}}]{1999ApJ...527..918W}
{Wyatt}, M.~C., {Dermott}, S.~F., {Telesco}, C.~M., {et~al.} 1999, \apj, 527,
  918

\end{thebibliography}


\begin{table*}
\centering
\caption{Opacity of the 0.32$''$ \textcolor{black}{NICI} Focal Plane Mask \citep{2013ApJ...779...80W}}
\begin{tabular}{lcccl}
\hline\hline 
Filters & Opacity (mag) & Uncertainty (mag) \\
\hline
$J$  & 6.37   & 0.12 \\
$H$ & 5.94   & 0.05 \\
$K_s$ & 5.70 & 0.09 \\
$CH_4S$ & 6.385 & 0.029 \\
$CH_4L$ & 6.204 & 0.048 \\
\hline
\label{tab:maf}
\end{tabular}
\end{table*}

\begin{table*}
\centering
\caption{NICI Observations of \prim .}
\begin{tabular}{llccl}
\hline\hline 
UT Date & Filters & N$_{frames}$ & Rotation~(\dg) & Comments\\
\hline
2012 April 7       & $J$                            &  66  &      73    &    Unsaturated star\\ 
2012 April 6       & $H$~$+$~$K$      &  74   &     81    &    Lightly saturated star\\
2009 January 14 & $CH_4$ short $+$ $CH_4$ long (4\%) &  49   &     24    &    Unsaturated star\\
2009 January 14 & $H$                            & 20   &      22    &   Saturated  star\\
\hline
\label{tab:obs}
\end{tabular}
\end{table*}

\begin{table*}
\centering
\caption{Ring Offset and Other Properties of the \prim\ Ring.}
\begin{tabular}{lccc}
\hline\hline 
Ring property & This work & \citet{2009AJ....137...53S} & \citet{2011ApJ...743L...6T}\\
\hline
$\Delta RA$ (mas)     & -7.4\pp0.8        &  -8\pp4\tablefootmark{a}  &      -6\pp4      \\ 
$\Delta DEC$ (mas)   & -15.0\pp1.1            &  -17\pp4\tablefootmark{a} &     -22\pp5     \\
$r$($''$)                 & 1.067\pp0.002  & 1.057\pp0.006  &     1.09\pp0.02 \\
PA (\dg)                    & 26.47\pp0.04    & 27.01\pp0.16   &      26.4\pp0.5      \\
Inclination (\dg)         & 76.0\pp0.07       & 75.9\pp0.6   &      76.7\pp0.5      \\
\hline
\label{tab:prevobs}
\end{tabular}
\tablefoot{
\tablefoottext{a}{Decomposed from reported value.}}
\end{table*}

\begin{table*}
\centering
\caption{Measurements of the Brightness Asymmetry of the \prim\ Ring.}
\begin{tabular}{cccc}
\hline\hline 
SW/NE & Filter & Radiation & Reference\\
\hline
0.85\pp0.02    &  $K_s$  & Scattered & This work  \\ 
0.78\pp0.03    &  $H$   & Scattered & This work  \\
0.93\pp0.02    &  $J$     & Scattered & This work  \\
0.92    &  0.53~\mic\   & Scattered  & \citet{2009AJ....137...53S}  \\
0.93    &  1~\mic\        & Scattered & \citet{2009AJ....137...53S}\\
0.6--1.0  &  1.10--2.22~\mic\        & Scattered & \citet{2008ApJ...673L.191D} \\ 
0.88    &  1.1~\mic\     & Scattered & \citet{1999ApJ...513L.127S,2005ApJ...618..385W} \\  
0.93    &  20.8~\mic\     & Thermal  &  \citet{2005ApJ...618..385W} \\   
0.96    &  24.5~\mic\     & Thermal  &  \citet{2005ApJ...618..385W} \\   
0.77    &  18.1~\mic\    & Thermal & \citet{2011AA...526A..34M}  \\
0.82    &  24.5~\mic\    & Thermal & \citet{2011AA...526A..34M}  \\
\hline
\label{tab:bright_asymm}
\end{tabular}
\end{table*}

\end{document}